\documentclass[twocolumn]{aa}  

\usepackage{graphicx}
\usepackage{txfonts}

\usepackage[colorlinks=true,urlcolor=blue,citecolor=blue,linkcolor=blue]{hyperref}

\begin{document}

    \title{Accounting for differential rotation in calculations of the Sun's angular momentum-loss rate}

   \titlerunning{Solar Wind Differential Rotation}
   \authorrunning{Finley \& Brun}

   \author{A. J. Finley\inst{1}
          \and
          A. S. Brun\inst{1}
          }

    \institute{Department of Astrophysics-AIM, University of Paris-Saclay and University of Paris, CEA, CNRS, Gif-sur-Yvette Cedex 91191, France \\ \email{adam.finley@cea.fr} }

   \date{Received December 30, 2022; accepted February 24, 2023}


  \abstract{Sun-like stars shed angular momentum due to the presence of magnetised stellar winds. Magnetohydrodynamic models have been successful in exploring the dependence of this ``wind-braking torque'' on various stellar properties, however the influence of surface differential rotation is largely unexplored. As the wind-braking torque depends on the rotation rate of the escaping wind, the inclusion of differential rotation should effectively modulate the angular momentum-loss rate based on the latitudinal variation of wind source regions.}{Here we aim to quantify the influence of surface differential rotation on the angular momentum-loss rate of the Sun, in comparison to the typical assumption of solid-body rotation.}{To do this, we exploit the dependence of the wind-braking torque on the effective rotation rate of the coronal magnetic field, which is known to be vitally important in magnetohydrodynamic models. This quantity is evaluated by tracing field lines through a Potential Field Source Surface (PFSS) model, driven by ADAPT-GONG magnetograms. The surface rotation rates of the open magnetic field lines are then used to construct an open-flux weighted rotation rate, from which the influence on the wind-braking torque can be estimated.}{During solar minima, the rotation rate of the corona decreases with respect to the typical solid-body rate (the Carrington rotation period is 25.4 days), as the sources of the solar wind are confined towards the slowly-rotating poles. With increasing activity, more solar wind emerges from the Sun's active latitudes which enforces a Carrington-like rotation. Coronal rotation often displays a north-south asymmetry driven by differences in active region emergence rates (and consequently latitudinal connectivity) in each hemisphere.}{The effect of differential rotation on the Sun's current wind-braking torque is limited. The solar wind-braking torque is $\sim 10-15\%$ lower during solar minimum, (compared with the typical solid body rate), and a few percent larger during solar maximum (as some field lines connect to more rapidly rotating equatorial latitudes). For more rapidly-rotating Sun-like stars, differential rotation may play a more significant role, depending on the configuration of the large-scale magnetic field.} 


   \maketitle
%

\section{Introduction}
The rotation periods of Sun-like stars slow systematically throughout the main-sequence \citep{skumanich1972time}, which can in some cases be used to determine the age of a star, or a population of stars \citep[a technique known as ``Gyrochronology''; see][]{barnes2007ages}. This is a consequence of magnetised stellar wind-braking, which enables the relatively feeble mass-loss rates of Sun-like stars to carry away significant amounts of angular momentum \citep{schatzman1962theory, weber1967angular, mestel1984angular, kawaler1988angular}. The wind-braking torque is often described as,
\begin{equation}
    \tau = \dot{M}\Omega_*\langle R_A\rangle^2,
    \label{tau}
\end{equation}
where $\dot{M}$ is the mass-loss rate of the wind, $\Omega_*$ is the rotation rate of the star, and $\langle R_A\rangle$ is the effective Alfv\'en radius of the wind, which essentially measures how far the magnetic field can exert a torque on the wind plasma before it becomes super-Alfv\'enic and effectively `disconnected' from the star \citep[see studies of][]{reville2015effect,finley2018dipquadoct}. 

The evolution of magnetism in Sun-like and low-mass stars, is largely constrained by systematic studies \citep[e.g.][and references therein]{see2019estimating} using the Zeeman broadening \citep[see review of][]{reiners2012observations} and Zeeman-Doppler imaging \citep{semel1989zeeman, donati2007magnetic} techniques. As is the evolution of their rotation periods, recovered from their long-term photometric variability \citep{curtis2019temporary, santos2021surface, rampalli2021three}. There are also a growing number of measured mass-loss rates derived from astrospheric Lyman-alpha absorption \citep{wood2021new}, slingshot prominences \citep{jardine2019slingshot}, and planetary transits \citep{vidotto2017exoplanets}. Though these observations are currently unable to fully-constrain the evolution of stellar mass-loss rates during the main sequence \citep{o2018solar}, and subsequently the effectiveness of stellar wind-braking \citep{brown2014metastable, matt2015mass, gondoin2017catastrophic, garraffo2018revolution, breimann2021statistical}. 

Studying the Sun directly bypasses many observational issues, given that the solar wind angular momentum flux can be measured in-situ \citep{lazarus1971observation,pizzo1983determination,marsch1984distribution,finley2019direct}. However the scale of the solar wind and the locality of the in-situ measurements often leads to issues of interpretation when computing a global wind-braking torque. In addition, all modern observations of the solar wind have been taken during a single epoch of its main-sequence lifetime \citep[see the review of][]{vidotto2021evolution}, whereas the typical wind-braking timescales for a star like the Sun is 10 - 100 Myrs. Using proxies for solar activity stored in natural archives, the longest reconstructions of solar activity are of the order of 10,000 years \citep{beer1998active, usoskin2017history, wu2018solar}. From this, the likely variations of the wind-braking torque can be inferred \citep{finley2019solar}, though this is still an order of magnitude away from being sensitive to the braking timescale of the Sun. It has also been suggested that the wind-braking of Sun-like stars begins to decrease significantly around the age of the Sun \citep{van2016weakened,metcalfe2016stellar,booth2017improved}. Evidence for this has begun to grow thanks to new asteroseismic observations \citep{hall2021weakened}, and models of the wind-braking torque for stars crossing the so called transition \citep{metcalfe2019understanding, metcalfe2022origin}. 

Despite the difficulties in its interpretation, a reliable assessment of the Sun's wind-braking torque has the potential to provide insight on the Sun's evolution and that of other Sun-like stars. In order to achieve this, there is a need to reconcile the wind-braking torques calculated using magnetohydrodynamic (MHD) models \citep[][]{reville2017global, finley2019effect}, and those calculated from in-situ measurements \citep[e.g.][]{finley2019direct}, which to-date disagree by a factor of a few. This may relate to the `open flux problem' \citep[][]{linker2017open, riley2019can}, in which the interplanetary magnetic field modelled by extrapolation from the observed photospheric magnetic field is systematically weaker than observed in-situ. Models that reproduce the observed value of the open magnetic flux in the solar wind tend to have better agreement with measurements of the in-situ angular momentum flux at 1au. Yet the angular momentum flux of the solar wind in the near-Sun environment, measured by Parker Solar Probe, has been shown to vary significantly from all model predictions, with tangential flows as strong as $\pm 50$ km/s \citep{kasper2019alfvenic}. 

Interestingly, when these large scale variations in the solar wind angular momentum flux are averaged over a solar rotation, they produce a wind-braking torque similar to that of the MHD models \citep[as shown for the first two encounters of Parker Solar Probe by][]{finley2020solar}. This suggests that the structure in the solar wind angular momentum flux develops in the low to middle corona (1 - 20 solar radii), either through the interaction of the magnetic field with rotation, or due to the interaction between neighbouring solar wind streams. An interaction-based mechanism for angular momentum redistribution appears to be supported by the prevalence of the fast solar wind carrying a negative (deflected) angular momentum flux, and the slow wind containing the more dominant net positive angular momentum flux \citep{finley2021contribution, verscharen2021angular}, i.e. removing angular momentum from the Sun. However the travel time from the solar surface to Parker Solar Probe (during encounters) is seemingly too short for wind-stream interactions to fully develop, perhaps indicating that the development of structure in the near-Sun environment is linked to the rotational state of the corona. Coronal rotation is also integral to the use of ballistic back-mapping for tracking solar wind plasma back to its source \citep[e.g.][]{macneil2022statistical}. 

This study makes a simplified assessment of the impact of differential rotation at the base of the solar wind on the resulting wind-braking torque. This is a first step towards future works examining the influence of more complex coronal rotation on the solar wind angular momentum flux. Section 2 sets out the data and methodology of the study, in which the Potential Field Source Surface (PFSS) model is used to rapidly realise the connectivity of the solar corona from a series of ADAPT-GONG magnetograms spanning one solar activity cycle (2007-2022). Section 3 presents the results of the inclusion of differential rotation in the calculation of the Sun's wind-braking torque. Finally, Section 4 puts our findings in context with current Heliospheric missions, and the winds of other Sun-like stars.

\section{Data and methodology}

\subsection{Rotation rate at the `coronal base'}

\begin{figure}
 \centering
  \includegraphics[trim=0cm 0cm 0cm 0cm, clip, width=0.45\textwidth]{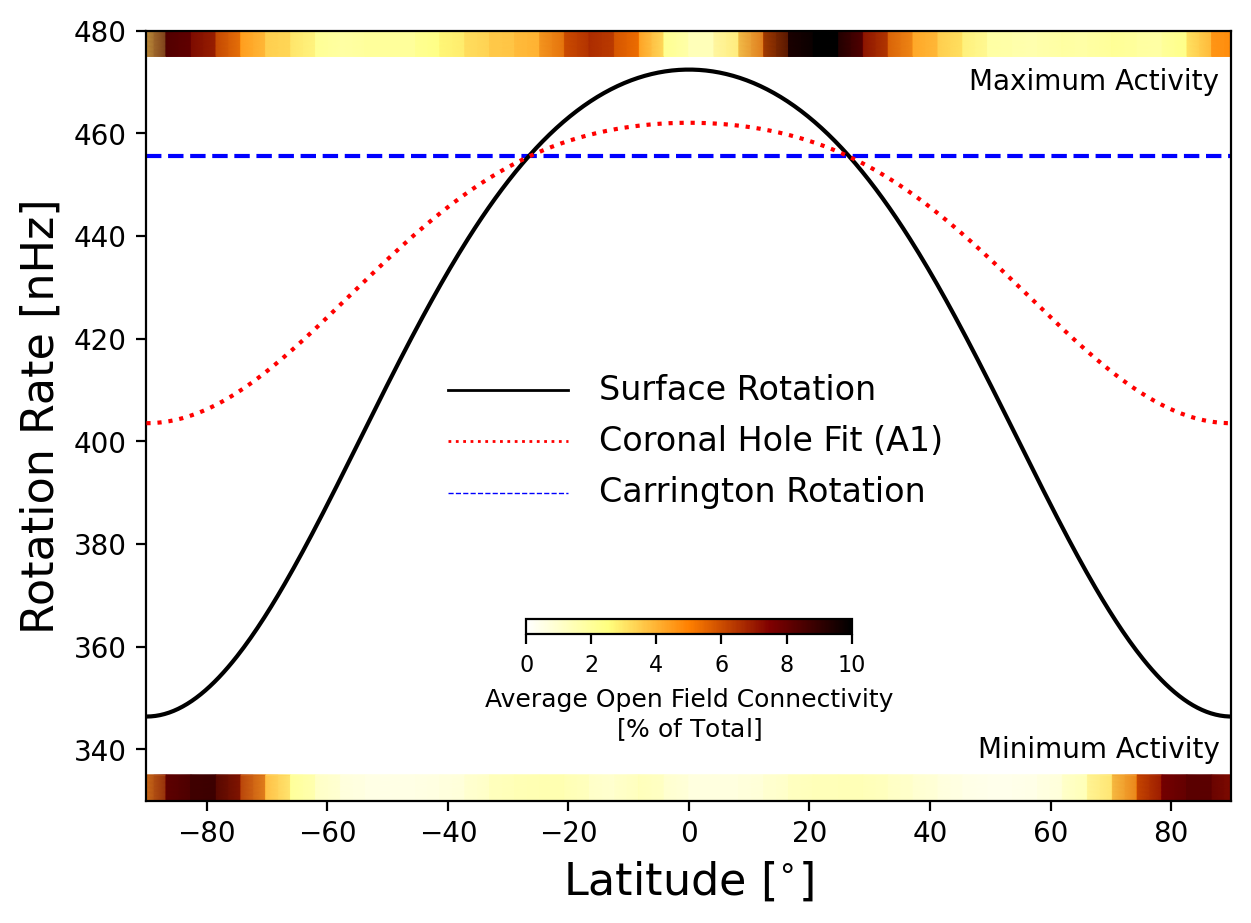}
   \caption{Rotation rate versus heliographic latitude. A typical solar differential rotation pattern is plotted with a solid black line. The rotation rate at $\theta\approx 26.5^{\circ}$, otherwise referred to as the Carrington rotation rate (of 25.4 days or $\sim 456$ nHz), is indicated with a blue dashed horizontal line. A less-extreme differential rotation pattern, taken from fitting the apparent motion of coronal holes in AIA-193A synoptic images (see Appendix \ref{AP_CH}), is plotted with a red dotted line. At the top and bottom of the figure, the latitudinal distribution of the source regions of the open magnetic field during the maximum activity and minimum activity periods of solar cycle 24, respectively, are indicated (further explored in Figure \ref{figure_latitudes}). During periods of low activity, the sources of the open magnetic field are confined to the slowly rotating poles. In periods of higher activity, the open field emerges more frequently from low-latitude features. During solar cycle 24, a north-south asymmetry is also observed during solar maximum.}
   \label{figure_DRprofile}
\end{figure}

\begin{table}
\caption{Rotation Rate Parameters.}
\label{table_rotation}
\centering
\begin{tabular}{c | c c c | c }     
\hline\hline
Profile & $\Omega_{eq}$ & $\alpha_2$ & $\alpha_4$ & Source\\
 & [nHz] & [nHz] & [nHz] & \\
\hline
   Surface & 472.6 & -73.9 & -52.1 & \citet{snodgrass1983magnetic}\\
   Coronal Hole & 463.0 & -26.7 & -33.1 & Appendix \ref{AP_CH}\\\hline
   Carrington & 455.7 & - & - & ---\\
\hline
\end{tabular}
\end{table}

The rotation of the solar surface is typically written in the form,
\begin{equation}
    \Omega_*(\theta) = \Omega_{eq}+\alpha_2\cos^2\theta+\alpha_4\cos^4\theta,
    \label{omega_*}
\end{equation}
where $\Omega_{eq}$ is the equatorial rotation rate, and the values of $\alpha_2$ and $\alpha_4$ describe the north-south symmetric differential rotation profile. Typical values describing the Sun's surface rotation rate are given in Table \ref{table_rotation}, taken from \citet{snodgrass1983magnetic}. Many studies have attempted to constrain the Sun's differential rotation profile \citep{newton1951sun,wilcox1970differential,howard1984rotation,beck2000comparison, lamb2017measurements,beljan2017solar,jha2021measurements}, however the profile from \citet{snodgrass1983magnetic} remains representative and is still frequently used in the literature. A schematic of this rotation rate profile versus latitude is shown in Figure \ref{figure_DRprofile}. Solar-like differential rotation has the equatorial plasma (and features embedded there) rotating faster than the polar regions. The equator rotates once around every 24.5 days, whereas the poles rotate once around every 33.4 days. Some strong magnetic features, however, appear to rotate around every 25.4 days (otherwise referred to as the Carrington rotation period). Whether this relates to the anchoring of magnetic field in the interior \citep[][]{gilman1983dynamically,miesch2008three, brun2004global, nelson2013buoyant, dikpati2021deciphering, kapyla2022transition, brun2022powering}, or the role of the near surface shear layer in sculpting the toroidal flux before emergence \citep[as discussed in][]{brandenburg2005case}, is unclear. Generally, magnetic features at the top of the convection zone are still subject to differential rotation, but this can be less that would be expected from the observed rate at the photosphere \citep[see][]{gigolashvili2013investigation}, and depends on their field strength and surface area \citep[][]{imada2018effect}. For some strong active regions, their observed shearing can be entirely independent of the global differential rotation pattern \citep[][]{yan2018successive}. 

Another diagnostic of rotation is the evolution of coronal holes in extreme ultraviolet (EUV) imagery, which appear dark as they are losing energy/mass directly to the (fast) solar wind. Previous authors have derived rotation rates from coronal holes in EUV \citep[see][for a study combining remote-sensing and in-situ observations]{heinemann2018three}. Again, these regions often appear to be less influenced by the surface differential rotation \citep{timothy1975structure, insley1995differential}, with some evidence for nearly rigid rotation in the low corona \citep{hiremath2013rotation}. Authors performing statistical studies of coronal hole rotation generally find a reduced amplitude of differential rotation \citep{bagashvili2017statistical, oghrapishvili2018study}. However it is unclear how these techniques are influenced by the latitudinal distribution of coronal holes, their appearance, evolution, and decay timescales, coupled with our limited ability to track them. In Appendix \ref{AP_CH}, a retrieval of the rotation profile from the apparent deformation of trans-equatorial coronal holes is shown through the comparison of multiple synoptic AIA-193$\AA$ charts. By maximising the coronal hole overlap from chart to chart, a reduced amplitude of differential rotation is recovered in each case. The average fit values are given in Table \ref{table_rotation}, and are comparable to those found in \citet{bagashvili2017statistical} and \citet{oghrapishvili2018study}. As the solar wind is accelerated over a large range of heights in the solar atmosphere, it is unclear what rotation rate should apply to the base of the wind (otherwise referred to as the coronal base). Therefore in Section \ref{results}, the assessment of the impact of differential rotation on calculations of the wind-torque is performed using both the surface and coronal hole motivated differential rotation profiles.

\subsection{Magnetic field extrapolation}

The use of PFSS models to infer the source locations of the solar wind (otherwise referred to as `connectivity') has become wide-spread due to the efficiency and simplicity of the model \citep{altschuler1969magnetic, schrijver2003photospheric}. Despite only taking information from the radial magnetic field at the surface, and having a single free parameter (the source surface radius, $R_{ss}$), this model has been shown to work well \citep{badman2020magnetic,panasenco2020exploring}, typically in advance of computing the more resource-intensive MHD models \citep[see comparison in][]{riley2006comparison}. For this study, our aim is to quantify the long-term variation in the field line mapping to different latitudes throughout the solar cycle. In this case, PFSS modelling represents a useful and reliable method to achieve this \citep[see similar work by][]{stansby2021active}. The PFSS magnetic field is constructed based on the following equations,
\begin{eqnarray}
B_r(r,\theta,\phi) = \sum_{l=1}^{\infty}\sum_{m=-l}^{l}\alpha_{lm}(r)Y_{lm}(\theta,\phi),\\
B_{\theta}(r,\theta,\phi) = \sum_{l=1}^{\infty}\sum_{m=-l}^{l}\beta_{lm}(r)Z_{lm}(\theta,\phi),\\
B_{\phi}(r,\theta,\phi) = \sum_{l=1}^{\infty}\sum_{m=-l}^{l}\beta_{lm}(r)X_{lm}(\theta,\phi),
\end{eqnarray}
where $r$ denotes radial distance from the origin, $\theta$ the latitude from the rotation pole, $\phi$ the Carrington longitude, and the typical $l$-degree and $m$-order spherical harmonic functions, using the legendre polynomial functions $P_{lm}(\cos\theta)$, are,
\begin{eqnarray}
Y_{lm}&=&c_{lm}P_{lm}(\cos\theta)e^{im\phi},\\ 
Z_{lm}&=&\frac{c_{lm}}{l+1} \frac{dP_{lm}(\cos\theta)}{d\theta} e^{im\phi}, \\ 
X_{lm}&=&\frac{c_{lm}}{l+1} P_{lm}(\cos\theta) \frac{im}{\sin\theta} e^{im\phi},
\end{eqnarray}
 with the normalisation of,
\begin{eqnarray}
c_{lm}=\sqrt{\frac{2l+1}{4\pi}\frac{(l-m)!}{(l+m)!}}.
\end{eqnarray}
In the PFSS model, the coefficients $\alpha_{lm}$ and $\beta_{lm}$ are given by, 
\begin{eqnarray}
\alpha_{lm}(r) = \epsilon_{lm}\frac{l(R_*/R_{ss})^{2l+1}(r/R_*)^{l-1}+(l+1)(r/R_*)^{-(l+2)}}{l(R_*/R_{ss})^{2l+1}+(l+1)},\\
\beta_{lm}(r) = (l+1)\epsilon_{lm}\frac{(R_*/R_{ss})^{2l+1}(r/R_*)^{l-1}+(r/R_*)^{-(l+2)}}{l(R_*/R_{ss})^{2l+1}+(l+1)},
\end{eqnarray}
where $\epsilon_{lm}$ represent the strength of each spherical harmonic mode. The $\epsilon_{lm}$ coefficients are extracted from the input magnetogram of the photospheric magnetic field by evaluating\footnote{To compute these coefficients the pySHTOOLS python package is used, which provides access to the Fortran-95 SHTOOLS library.},
\begin{eqnarray}
    \epsilon_{lm}=\frac{1}{c_{lm}}\int_{\phi}\int_{\theta} B_r(\theta,\phi)P_{lm}(\cos\theta)\cos(m\phi)\sin\theta d\theta d\phi.
\end{eqnarray}

In this study, the PFSS model is driven by spherical harmonic decomposition of the ADAPT-GONG magnetograms\footnote{Data accessed Feb 2022: https://nso.edu/data/nisp-data/adapt-maps/} \citep{arge2010air}. These magnetograms are produced using a combination of data assimilation and forward modelling, which accounts for the effects of differential rotation, meridional circulation, and diffusion on older observations \citep[further discussed in][]{hickmann2015data}. In general, the Sun's polar magnetic fields are difficult to capture due to their proximity to the limb and weak line-of-sight strength. The ADAPT-GONG magnetograms leverage the underlying flux transport model to reproduce the time evolution of the polar fields based on data assimilated at lower latitudes. This has been shown to remain consistent with direct observations of the polar field \citep{arge2011improving}, and so reduces the potential variation in the reconstructed coronal magnetic field structure caused by the varying visibility of the Sun's poles. Magnetograms are taken at a monthly cadence from Jan 2007 to Feb 2022 ($\sim200$ Carrington Rotations), always using the first realisation of the magnetogram. To reduce computational cost, the magnetic field is reconstructed up to a spherical harmonic degree of $l_{max}\approx30$ using $R_{ss}=2.5R_{\odot}$ on an equally-spaced grid of $r\times\theta\times\phi$ resolution of $12 \times 92 \times 184$ points. This set-up is found to reliably recovers structures based on the resolution of the ADAPT-GONG magnetogram inputs.

\subsection{Effective rotation rate}

\begin{figure}
 \centering
  \includegraphics[trim=0cm 0cm 0cm 0cm, clip, width=0.5\textwidth]{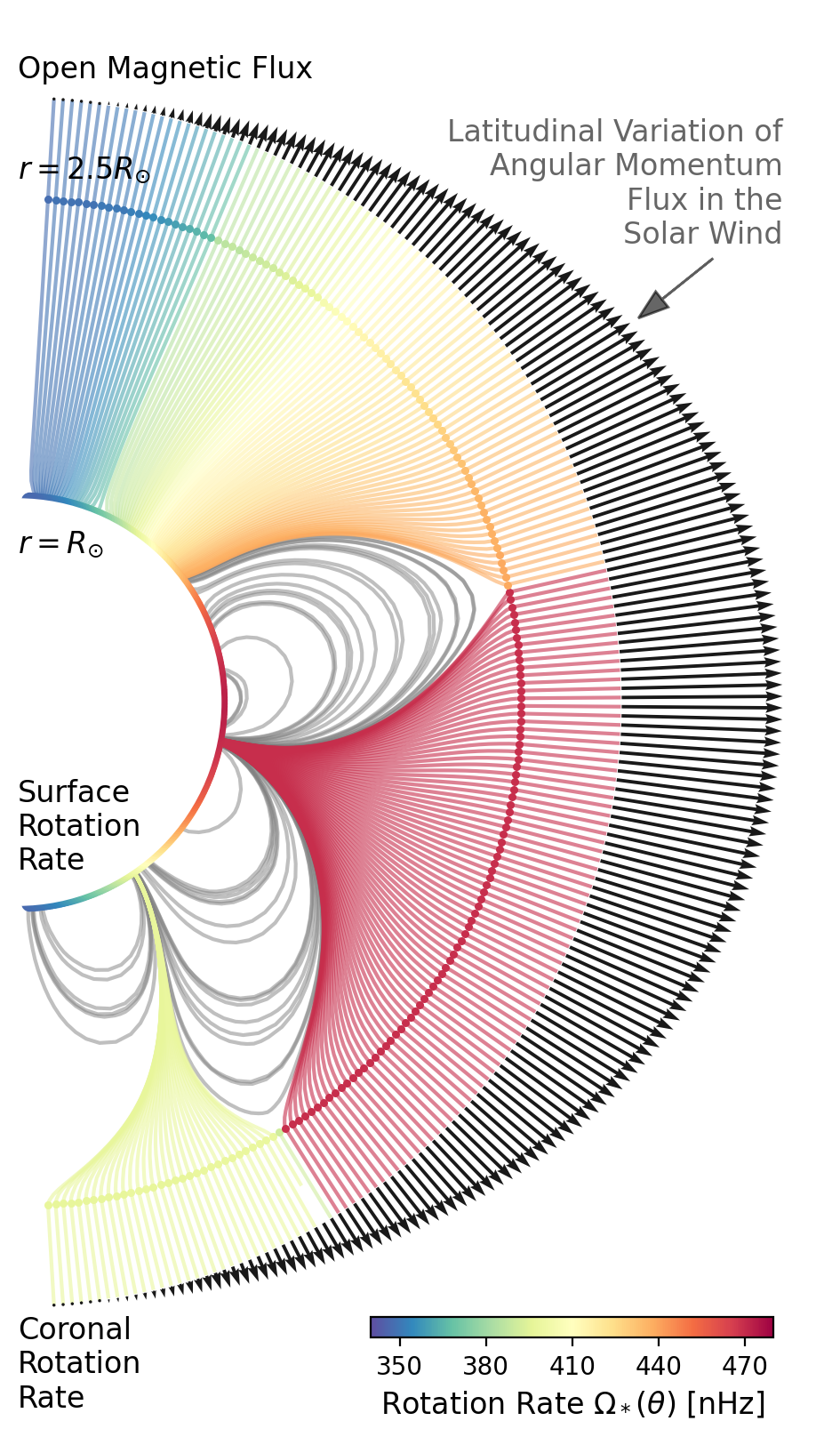}
   \caption{Schematic depiction of the methodology used in this study. The surface rotation rate is extrapolated into the corona via a PFSS model driven by ADAPT-GONG magnetograms. In three dimensions this produces a longitude-latitude grid of rotation rates at the height of the source surface ($R_{ss}=2.5R_{\odot}$), the height at which the coronal magnetic field opens into the solar wind. The open flux weighted rotation rate is then calculated from the longitude-latitude grid of rotation rates, however the angular momentum flux in the solar wind is expected to vary with distance from the rotation axis. Therefore a factor of $\sin\theta$ is introduced into open flux weighted averaging process, giving additional weight to the solar wind close to the equator (represented by the magnitude of the black arrows). }
   \label{figure_schematic}
\end{figure}

\begin{figure*}
 \centering
  \includegraphics[trim=0cm 0cm 0cm 0cm, clip, width=\textwidth]{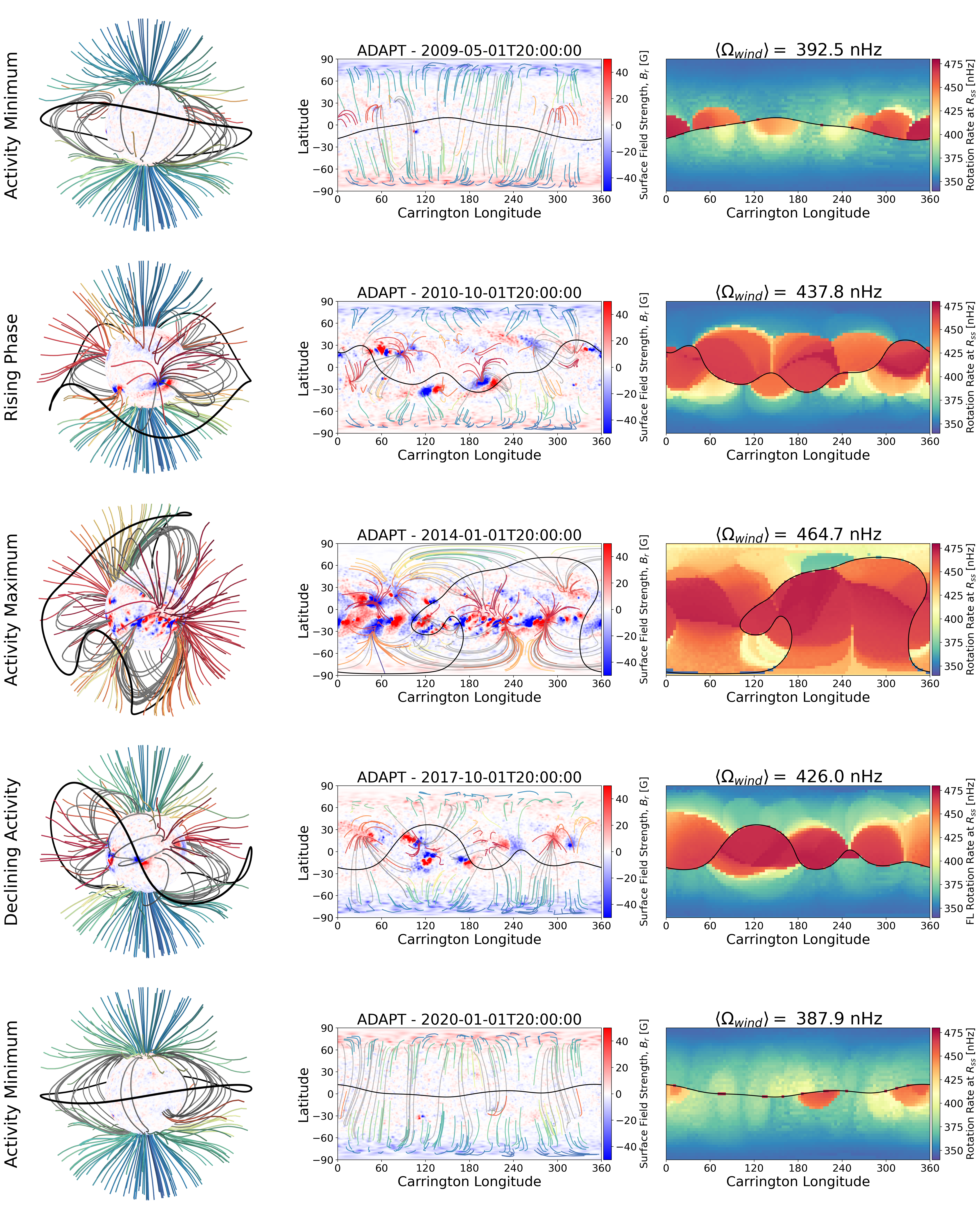}
   \caption{Summary of five models from different phases of solar cycle 24. The first column shows 3D renderings of the PFSS models, with open magnetic field lines coloured by surface rotation rate (see Figure \ref{figure_DRprofile}). Closed magnetic field lines are shown in grey. The solar surface is coloured with red and blue representing the radial magnetic field used in each PFSS model, and the location of the Heliospheric Current Sheet (HCS) is indicated in black at the source surface. The center column displays the same information projected on a latitude-longitude grid. The final column shows the effective rotation rate of field lines at the source surface ($R_{ss}=2.5R_{\odot}$). These values are acquired by tracing field lines down to the surface, and returning the value of the surface rotation rate (this is the same as for the colour on the field line renderings). The mean value of the effective rotation rate $\langle \Omega_{wind}\rangle$ is listed with each model (the Carrington rate is $\sim 456$ nHz).}
   \label{figure_3Dplot}
\end{figure*}

In ideal MHD models of solar/stellar wind, the footpoints of the magnetic field should remain anchored to the surface (or inner boundary condition). This requires, for a perfectly conducting rigidly rotating boundary with a frozen-in magnetic field, that the electric field at the surface in the rotating frame be zero. In the case of a steady state solution, i.e. axisymmetric or rigid rotation, this condition produces a scalar quantity which is constant along magnetic field lines, the effective rotation rate \citep{mestel1968magnetic, sakurai1985magnetic}. This quantity in CGS units is,
\begin{equation}
    \Omega_{eff} = \frac{1}{r\sin\theta}\bigg(v_{\phi}-\frac{B B_{\phi}}{4\pi\rho v}\bigg),
    \label{omega_eff}
\end{equation}
where $r$ is the radius, $\theta$ is the latitude from the rotation pole, $v$ is the fluid velocity, $B$ is the magnetic field vector, $\rho$ is the fluid density, and quantities with subscript $\phi$ are taken in the azimuthal direction. This quantity is typically set at the lower boundary to enforce a given surface rotation rate $\Omega_{eff}=\Omega_*$ \citep[e.g.][]{zanni2009mhd}. 

In the recent work of \citet{ireland2022effect}, the value of $\Omega_{eff}$ was allowed to vary in latitude in order to model the influence of differential rotation on the wind-braking torques from their 2.5D stellar wind models \citep[see also][]{pinto2021solar}. This has the effect of anchoring the field lines at different latitudes to different rotation rates. As this model is axisymmetric, the effective rotation rate remained constant along magnetic field lines (the conservation of this quantity is also used to validate the performance of the numerical methods). The authors tested varying degrees of solar-like differential rotation, whilst also altering the stellar magnetic field strength. The resulting wind-braking torques were shown to be well-behaved, and described by a correction factor that accounted for the implicit change in the rotation rate $\langle\Omega_{wind}\rangle$ of the simulation in comparison to solid-body $\Omega_*$. As the stellar magnetic field in this case was dipolar, it was shown that the rotation rate of the wind scaled with the latitude of the last open magnetic field line. 

In the case of a 3D non-axisymmetric magnetic field with differential rotation, shearing in the corona creates a time-dependant solution which is sensitive to the degree of non-axisymmetry in the magnetic field, and the contrast in rotation rate between the footpoints of closed coronal loops. A steady-state solution is unlikely to be reached in this case. Irrespective of this, the rotation rate of the open magnetic field will likely still be strongly influenced by the anchoring speed of the footpoints. Given the relatively slow rotation of the Sun, in this study these interactions are assumed to be weak, such that the effective rotation rate is conserved along each field line. This allows for the effective rotation rate to be propagated into the corona.


\begin{figure*}
 \centering
  \includegraphics[trim=0cm 0cm 0cm 0cm, clip, width=\textwidth]{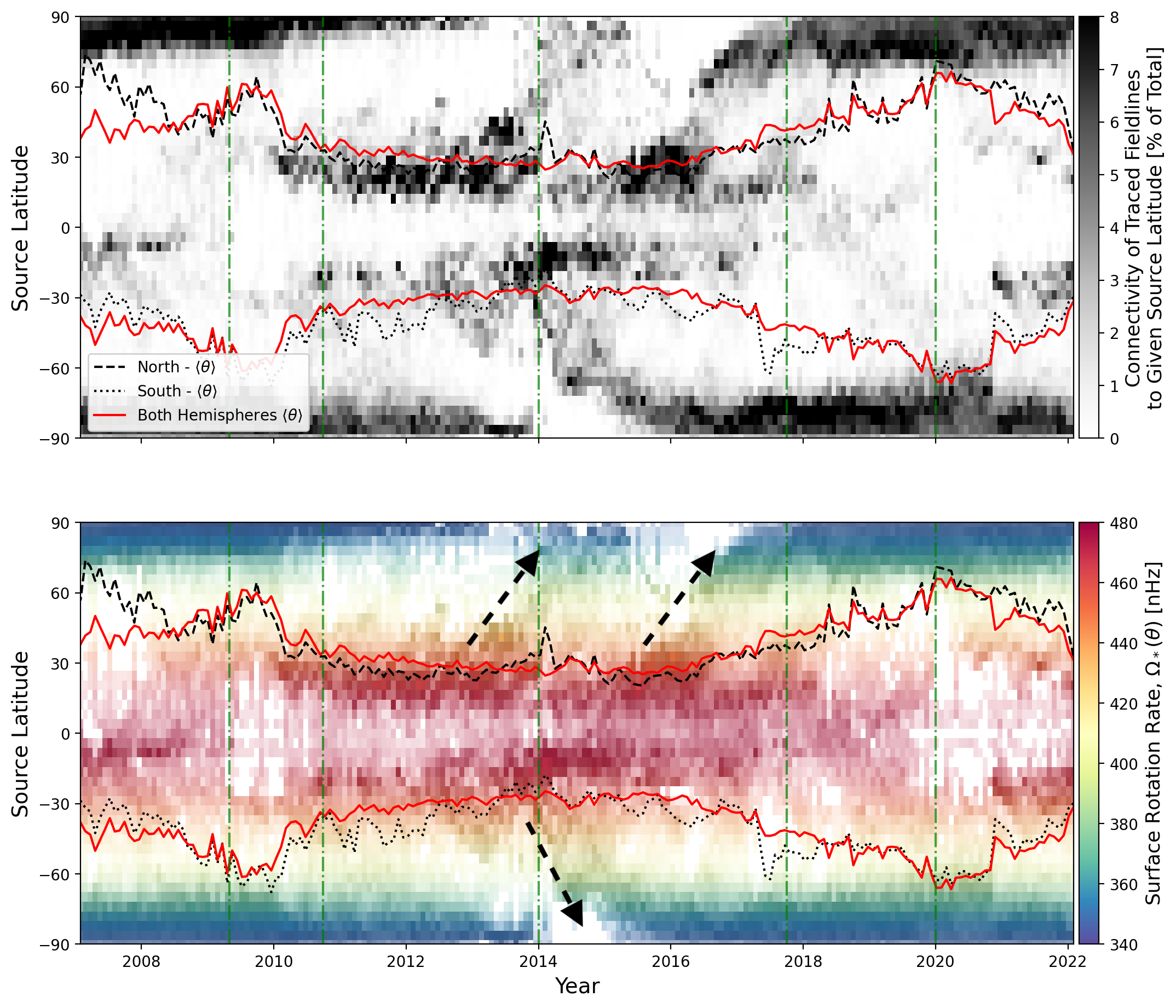}
   \caption{Histogram of the source latitudes of magnetic field lines traced down from the source surface for each magnetogram in our time-series. The top panel indicates, for each model, the percentage of field lines that connect to a given latitude (bin width $\sim2^{\circ}$) from a homogeneous sampling of the source surface. The bottom panel blends this information with that of the rotation rate at those latitudes, i.e. the more vivid the colours the larger the fraction of field lines connecting to that latitude (as in the panel above). Arrows indicate the major pole-ward surges of magnetic flux during solar cycle 24. The snapshots shown in Figure \ref{figure_3Dplot} are identified with green vertical dot-dashed lines in both panels. The mean latitude of connectivity (including a factor of $\sin\theta$ to match $\langle \Omega_{wind}\rangle$) is shown with red solid lines. The same calculation is repeated for the northern and southern hemispheres individually, plotted in black dashed and dotted lines respectively; highlighting the degree of asymmetry.}
   \label{figure_latitudes}
\end{figure*}

In addition to the significance of the field line footpoint rotation, field lines closer to the equator will carry a larger angular momentum-flux \citep[e.g.][]{keppens1999numerical}, due to the geometrical lever arm from the rotation axis. Thus the rotation rate of field lines nearer the equator will have a stronger influence on the mean rotation rate that is needed to describe the wind-braking torque with equation (\ref{tau}). The mean rotation rate of the wind, is therefore calculated via an open-flux weighted average in the magnetically-open corona including a $\sin\theta$ dependence,
\begin{equation}
    \langle\Omega_{wind}\rangle=\frac{\oint_{A}\Omega(r,\theta)\sin\theta|{\pmb B}\cdot d{\pmb A}|}{\oint_{A}\sin\theta|{\pmb B}\cdot d{\pmb A}|},
    \label{omega_wind}
\end{equation}
where $\Omega(r,\theta) = \Omega_*(R_*,\theta_*)$ is the value of the surface differential rotation rate mapped along the magnetic field from $(R_*,\theta_*)$ to $(r,\theta)$, and the closed integral over the area $A$ of the magnetic field vector $\pmb B$ returns the unsigned magnetic flux in the wind. The radius $r$ should therefore be larger than the last closed magnetic field loop. The dependence of the angular momentum flux on latitude from the rotation axis is further discussed in \citet[][]{finley2019direct}.


A schematic depiction of this calculation is shown in Figure \ref{figure_schematic}. The technique described here could also be used to correct the wind-braking torques from MHD wind simulations performed with solid-body rotation. The form of $\langle\Omega_{wind}\rangle$ is motivated by insights from the scaling of the wind-braking torque in \citet{ireland2022effect}. It is left for future work to truly validate this relation. In addition, throughout this work, the effective rotation rate of the magnetic field lines and that of the solar wind are assumed to be interchangeable, however this is an oversimplification of equation (\ref{omega_eff}). The development of stress in the magnetic field can also modify the rotation rate of the wind. This does not affect the calculation of $\langle\Omega_{wind}\rangle$, but caution should be used in the interpretation of the extrapolated coronal rotation rates. To self-consistently model coronal rotation,  time-dependent simulations which are continuously driven, such as magnetofrictional models \citep[e.g.][]{yeates2013coronal,hoeksema2020coronal}, may be better suited (though more computationally expensive).

\section{Results}\label{results}

\subsection{Global connectivity and rotation rate}

Figure \ref{figure_3Dplot} displays PFSS models for five different ADAPT-GONG magnetograms (2009-06-01, 2010-10-01, 2014-01-01, 2017-10-01, and 2020-01-01), each representing a different phase of solar cycle 24. 3D renderings are shown in the first column with magnetic field lines coloured by the surface rotation rate, following equation (\ref{omega_*}). Closed field line are coloured grey. This information is re-projected onto a latitude-longitude grid in the center column. The final column shows a latitude-longitude map of the effective rotation rate at $2.5R_{\odot}$ created by tracing field lines down from an equally-spaced grid of $48\times 96$ points at the source surface ($r=R_{ss}$) to sample the corresponding surface rotation rate (see Figure \ref{figure_schematic}). From this map, equation (\ref{omega_wind}) is evaluated and the value of $\langle\Omega_{wind}\rangle$ is noted with each model. This calculation is repeated for each magnetogram in our sample. 

Figure \ref{figure_3Dplot} shows some clear trends during the solar cycle. At minima of solar activity (top and bottom rows), the open magnetic field emerges primarily from the slowly rotating polar coronal holes. There are some additional contributions of open magnetic field from small active regions or equator-ward coronal holes which create pockets of fast rotation near to the Heliospheric Current Sheet (HCS). In more active phases (middle rows), the HCS becomes increasingly warped by underlying active regions, and equatorial coronal holes. These low-latitude sources of the solar wind increases the proportion of fast rotation in the corona. At solar maximum, the dipole axis is completely tilted, closing off most of the field at the poles, leading to faster rotation throughout the entire corona. There are subtle differences between the rising and decay phases of the cycle, with surges of magnetic flux towards the poles during the decay phase leading to more extended polar coronal holes in latitude and accordingly coronal rotation rates that are slightly elevated with respect to the rising phase. Throughout the cycle, the degree of warping of the HCS appears an indirect indicator of the rotational state of the corona, with deviation from a perfectly flat HCS in the equator most-likely due to source regions at low-latitude, anchored at more rapidly rotating latitudes.

The variation of solar wind source regions during solar cycle 24 is shown more clearly in Figure \ref{figure_latitudes}. Histograms of the open magnetic field footpoint latitudes are shown for each model in our time-series of $\sim 200$ magnetograms. Open field lines are traced down from the source surface with seeds distributed homogeneously over all latitudes and longitudes. Values in the top panel are given as a percentage of field lines from the total seeds that connected to a given latitude bin (width $\sim 2^{\circ}$), irrespective of longitude. During solar minimum, the open magnetic field is mostly confined to the rotational poles, which is in contrast to solar maximum where the majority of the open magnetic field emerges from lower-latitudes. The bottom panel of Figure \ref{figure_latitudes} is similar, but with result coloured by the surface rotation rate. The presence of multiple pole-ward rushes in magnetic flux is identifiable in the field line connectivity (highlighted with dashed arrows). 

\begin{figure}
 \centering
  \includegraphics[trim=0cm 0cm 0cm 0cm, clip, width=0.49\textwidth]{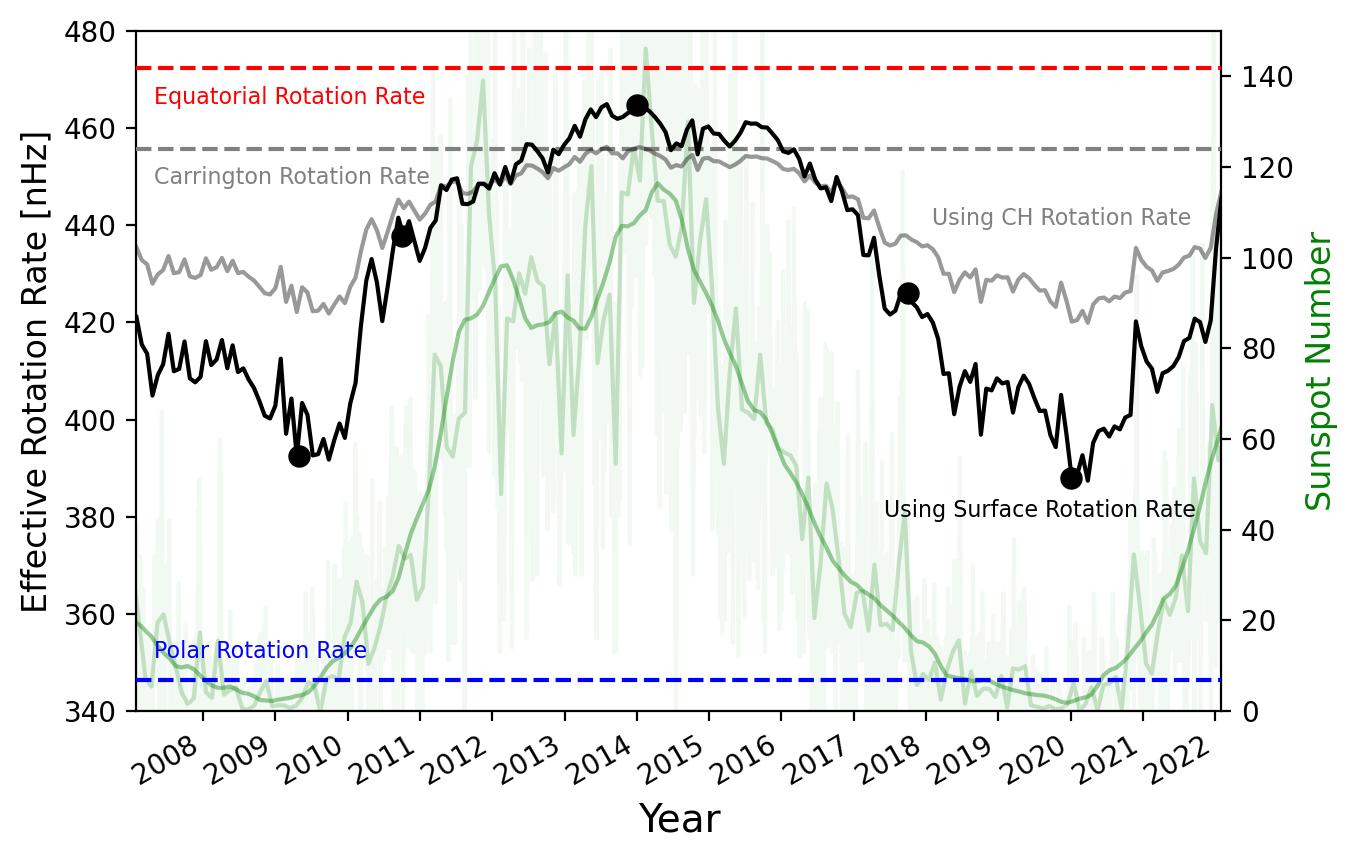}
   \caption{Effective rotation rate versus solar cycle, calculated with equation (\ref{omega_wind}). Solid black line represents the ADAPT-GONG magnetograms using $R_{ss}=2.5R_{\odot}$, and the observed surface rotation rate. Dashed lines indicate the equatorial, polar and Carrington rotation rate. The solid grey line instead uses the less-extreme coronal hole rotation profile (see Appendix \ref{AP_CH}). PFSS models shown in Figure \ref{figure_3Dplot} are highlighted with black circles. The daily, monthly, and monthly-smoothed sunspot number from the Sunspot Index and Long-term Solar Observations (SILSO) are displayed in the background of the figure with a solid green lines of varying opacity.}
   \label{figure_effRotation}
\end{figure}

The value of $\langle\Omega_{wind}\rangle$ calculated from equation (\ref{omega_wind}) using the surface rotation rate and the coronal hole rotation rate is shown in Figure \ref{figure_effRotation} versus time, along with the sunspot number. During solar minimum, $\langle\Omega_{wind}\rangle$ is smaller than the Carrington rate, however does not reach the polar rotation rate due to the volume-filling patches of low-latitude connectivity. These regions are close to the equator and are therefore more heavily-weighted by the additional $\sin\theta$ dependence in equation (\ref{omega_wind}). With increasing solar activity, the value of $\langle\Omega_{wind}\rangle$ increases to be close to, but slightly larger than, the Carrington rotation rate, as more open magnetic field is emerging from the active latitudes. Naturally, this leads to a correlation between $\langle\Omega_{wind}\rangle$ and the amount of open magnetic flux in the wind. As typically an increase in open magnetic flux results from increased flux emergence around the active latitude (which are evidently rotating faster than the poles). 

\begin{figure}
 \centering
  \includegraphics[trim=0cm 0cm 0cm 0cm, clip, width=0.49\textwidth]{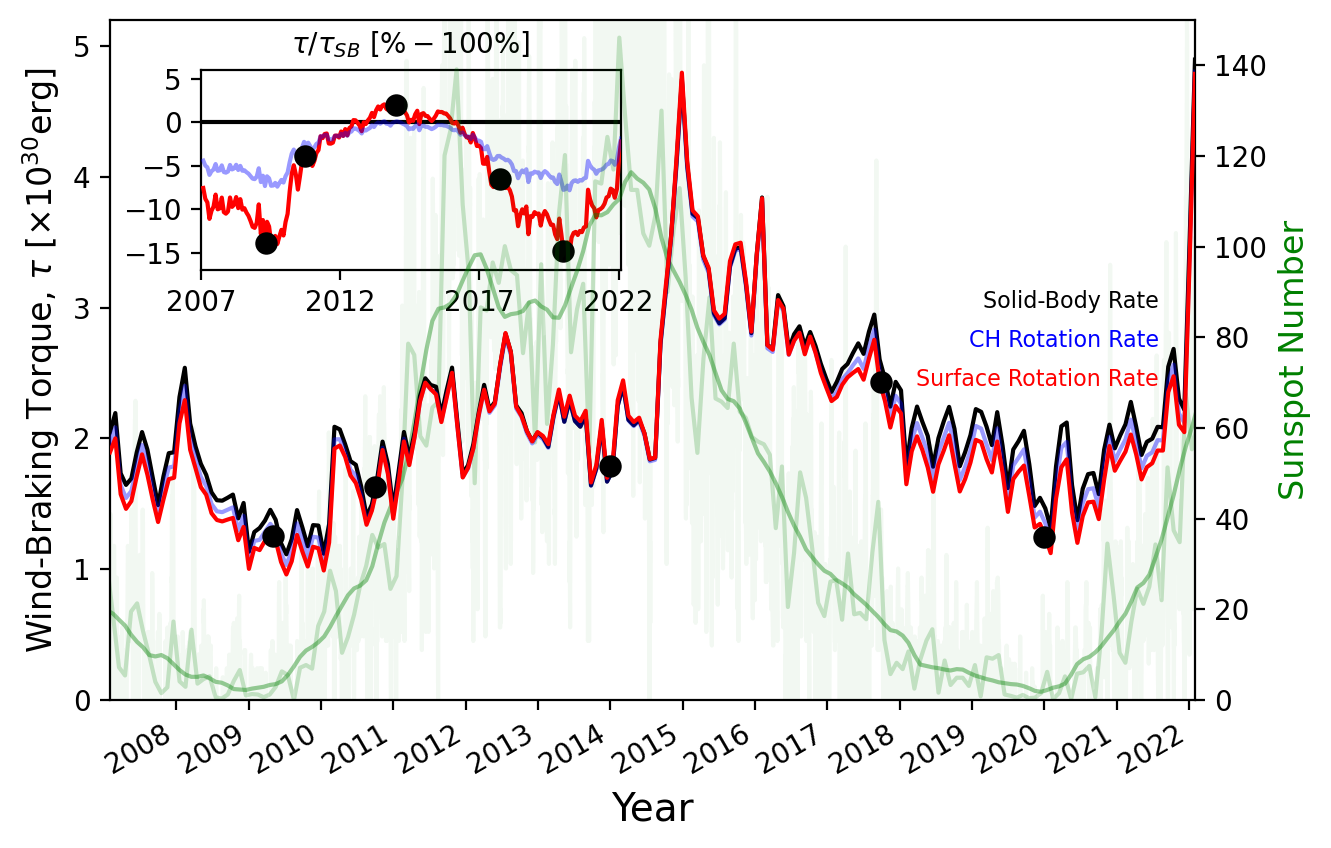}
   \caption{Same as Figure \ref{figure_effRotation}, but now showing the expected change in wind-braking torque between solid body (SB) and differentially rotating (DR) models on the vertical axis. }
   \label{figure_torque}
\end{figure}

\begin{figure*}
 \centering
  \includegraphics[trim=0cm 0cm 0cm 0cm, clip, width=\textwidth]{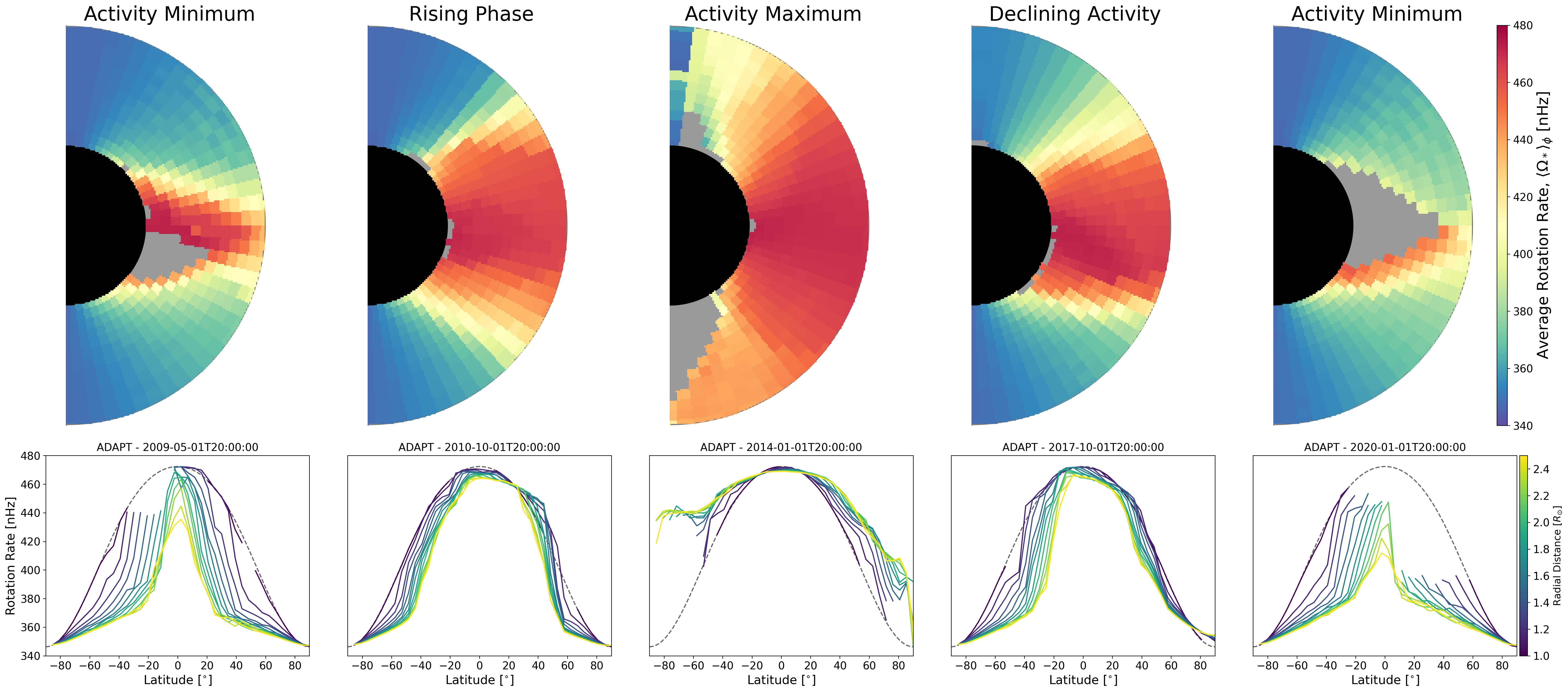}
   \caption{Azimuthally-averaged rotation rates from the PFSS models shown in Figure \ref{figure_3Dplot}. Averaging is peformed on the open field regions, grey regions indicate closed field at all longitudes (hence no value returned). The limited number of open magnetic field lines over the north pole during the `Activity Maximum' model, lead to some oddity in the north-most latitude bin that should be disregarded. The lower panels show the latitudinal profiles coloured by radial distance from the surface (yellow being the furthest). The rising and declining phases show a clear north-south asymmetry which is also observed in the systematic differences in $\langle\theta\rangle$ between the two hemispheres in Figure \ref{figure_latitudes}. } 
   \label{figure_1_AZAV}
\end{figure*}

The open magnetic flux is also increased by lowering the source surface height. The results presented so far use a fixed value of the source surface radius ($R_{ss}=2.5R_{\odot}$), however this value is likely to vary during the solar cycle \citep{arden2014breathing, pinto2011coupling, perri2018simulations, hazra2021modeling}. Our analysis is repeated in Appendix \ref{AP_RSS} with source surface radii of 2 $R_{\odot}$ and 3 $R_{\odot}$, to investigate the dependence of $\langle\Omega_{wind}\rangle$ on the source surface height (see Figure \ref{AP_RSS}). As expected, the smaller the source surface the more higher-order magnetic field is opened which increases the open magnetic flux. Once again, this shifts connectivity towards smaller active regions. During solar minimum, this means a smaller source surface can more-easily connect to the faster equator-ward latitudes (increasing $\langle\Omega_{wind}\rangle$ by around $20$ nHz), with the opposite effect for larger source surfaces (decreasing by $10$ nHz). This is further detailed in Appendix \ref{AP_RSS}.

\subsection{Solar wind angular momentum-loss rate}

MHD models have been used to explore the dependence of the wind-braking torque on various configurations of the coronal magnetic field, under different coronal heating scenarios, and rotation rates \citep{matt2012magnetic, reville2015effect,pantolmos2017magnetic,finley2018dipquadoct,hazra2021modeling, ireland2022effect}. In the slowly rotating regime, where centrifugal effects on the wind acceleration are negligible, the mass-loss rate and Alfv\'en radius are unaffected by the inclusion of differential rotation. This results in a linear dependence between the wind-braking torque and the effective rotation rate, as in equation (\ref{tau}). Provided that equation (\ref{omega_wind}) is a reasonable approximation for the rotation rate of the wind, changes to the wind-braking torque $\tau_{DR}$ are then given by,
\begin{equation}
    \tau_{DR} = \frac{\langle\Omega_{wind}\rangle}{\Omega_*}\tau_{SB},
    \label{tau_DR}
\end{equation}
where $\tau_{SB}$ is the wind-braking torque calculated using the solid-body rotation value $\Omega_*$ (which is taken to be the Carrington rotation rate). For the Sun, $\tau_{SB}$ has been calculated by many authors using a variety of models and semi-analytic relations. In general, the wind-braking torque is largest during periods of increased solar activity as the amount of open magnetic flux in the solar wind is increased. Here, the semi-analytic relation for the solar wind-braking torque from \citet{finley2019solar} is adopted. This relation is derived from a parameter study of 2.5D MHD simulations in \citet{finley2017dipquad,finley2018dipquadoct}. The wind-braking torque is given by,
\begin{eqnarray}
    \tau_{SB} =(2.3\times10^{30}[\text{erg}])\bigg(\frac{\dot M}{1.1\times 10^{12} [\text{g/s}]}\bigg)^{0.26} \nonumber \\
    \times \bigg(\frac{\phi_{open}}{8.0\times 10^{22}[\text{Mx}]}\bigg)^{1.48},
    \label{open_torque}
\end{eqnarray}
where $\dot{M}$ is the solar mass-loss rate, and $\phi_{open}$ is the open magnetic flux in the solar wind. Both of these variables are estimated from in-situ measurements of the solar wind from the \textit{Wind} spacecraft, as done in \citet{finley2019solar}. In-situ measurements from the equatorial solar wind are averaged on the timescale of a Carrington rotation ($\sim 27$ days as viewed from Earth), in order to remove longitudinal structures. Latitudinal variations in the mean mass flux and magnetic flux are assumed to be small at 1au, such that the averaged equatorial values can be used to create global estimates of $\dot{M}$, and $\phi_{open}$. From these values, the solar wind-braking torque computed with equation (\ref{open_torque}) is plotted in Figure \ref{figure_torque}. Applying the value of $\langle\Omega_{wind}\rangle/\Omega_*$ to the solar wind-braking torque, produces the corrected torque $\tau_{DR}$ for the surface and coronal hole profiles (plotted in Figure \ref{figure_torque}  with red and blue lines respectively).

The percentage change in the wind-braking torque ($\tau_{DR}/\tau_{SB}\times100\%$) during the solar cycle varies from 10-15$\%$ during solar minima, to a few percent at solar maximum (see inset of Figure \ref{figure_torque}). In times of increased solar activity, the equatorial solar wind can emerge from sources closer to the equator than $\theta\approx \pm 26.5^{\circ}$ (e.g. the Carrington rate latitude). This results in a more rapidly rotating corona than the typical solid body value, and hence a slightly larger wind-braking torque. However, as previously discussed, equatorial coronal holes visible in EUV imagery do not always show this rapid rotation, and so in reality, the equatorial rotation rate may be closer to that of the coronal hole motivated profile. Using this rotation profile instead of the typical differential rotation profile reduces the effect at solar minima (to around 5$\%$) and results in a negligible change during solar maximum. In either case, the effect of differential rotation tends to reinforce the pre-existing variation of the Sun's angular momentum-loss rate during the solar cycle. With the strongest influence of differential rotation occurring when the solar wind-braking torque is smallest, the overall impact on the long-term angular momentum-loss rate is minimised. 

The weak dependence of the wind-braking torque on the Sun's differential rotation profile is easily explained by considering the extreme values that $\langle\Omega_{wind}\rangle$ could take. These being the polar and equatorial rotation rates, i.e. all the wind rotates either at the slowest or fastest possible rotation rate. In this case, taking the solid-body rotation rate to be that of the poles, around 33.4 days, will produce a solar wind-braking torque that is $\sim 24\%$ smaller than using the Carrington rotation rate. Similarly, by using the equatorial rotation rate of 24.5 days the wind-braking torque increases by 4$\%$. Given the small variation in wind-braking torques between these maximum and minimum values of the effective rotation rate, the use of solid-body rotation in steady-state MHD models of the solar wind appears to be justified to first order. 

\begin{figure*}
 \centering
  \includegraphics[trim=0cm 0cm 0cm 0cm, clip, width=\textwidth]{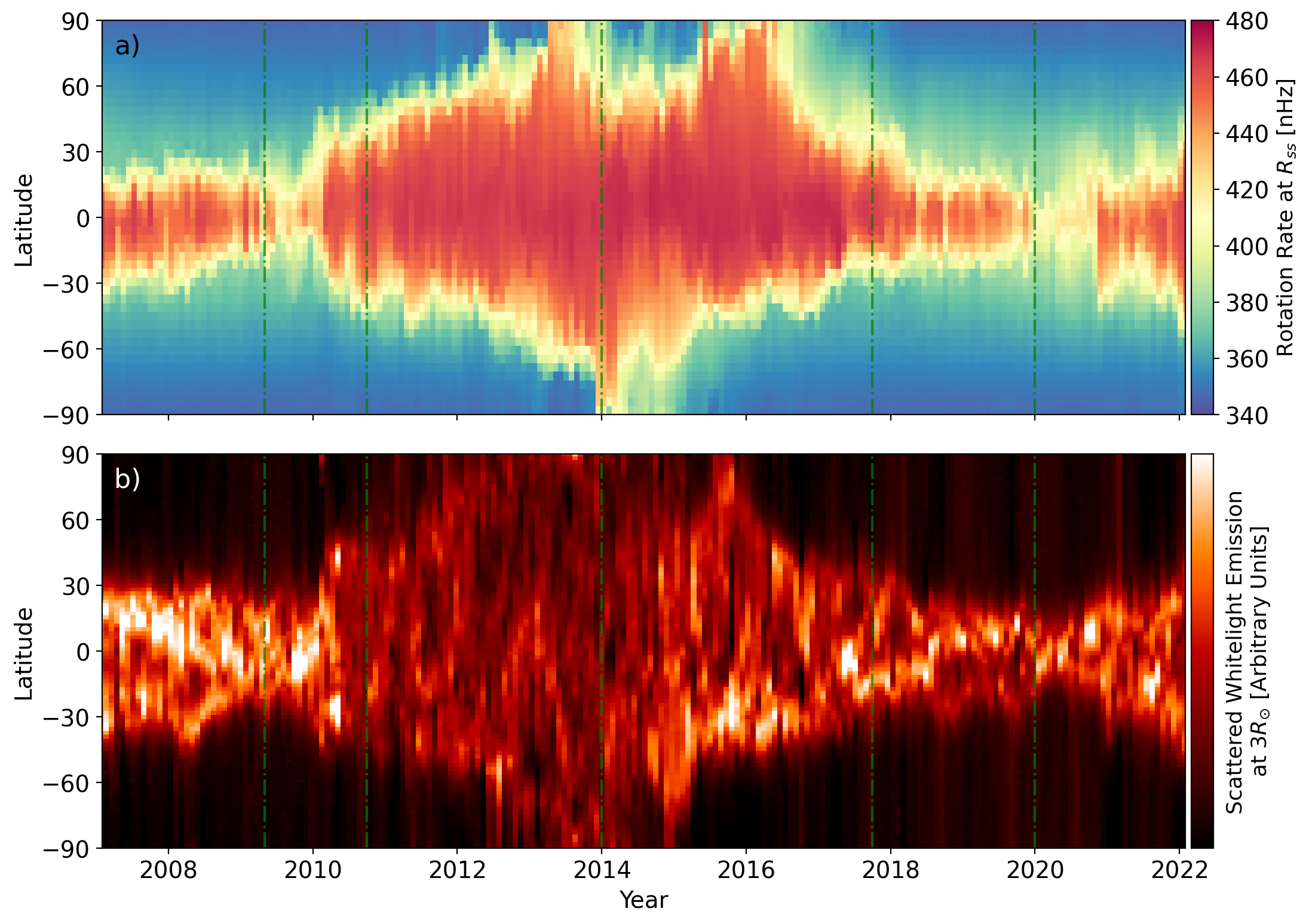}
   \caption{The potential bias from detecting coronal rotation from white light streamer structures. Panel a) depicts the azimuthally-averaged rotation profile versus latitude at the source surface ($R_{ss}=2.5R_{\odot}$) throughout the timeseries of magnetograms. Panel b) charts the average brightness of scattered white light from streamers at three solar radii observed by LASCO-C2 onboard Solar and Heliospheric Observatory. Streamers are confined to the equator near minimum activity and spread to all latitude during maximum. Given that the apparent motion of white light streamers is used to measure coronal rotation, this technique may be biased towards recovering the Carrington rotation rate at most available latitudes. } 
   \label{figure_lasco}
\end{figure*}

\section{Discussion}

\subsection{Asymmetric Rotation of the Corona}\label{ass}

The mean latitude of connectivity during the solar cycle (weighted by a factor of $\sin\theta$ to be consistent with the definition of $\langle\Omega_{wind}\rangle$) is plotted in Figure \ref{figure_latitudes} with symmetric solid red lines in each hemisphere. Performing this calculation independently for the northern and southern hemispheres produces the black dashed and dotted lines respectively. Deviation from the symmetric solid red lines indicates asymmetry between the two hemispheres. 

As solar cycle 24 progresses, active regions appear in the north and south following the typical butterfly-pattern \citep[e.g.][]{hathaway2015solar}. At first, active latitudes in the north are more frequently sources of the solar wind than in the south, which pulls the mean latitude closer to the equator in the north than in the south. This leads to an asymmetric increase of coronal rotation in the northen hemisphere. After a pole-ward surge in the northern hemisphere (starting in 2013), the connectivity begins to favour the southern active latitudes. This briefly reverses the asymmetry in the coronal rotation, producing a faster southern hemisphere. A pole-ward surge in the southern hemisphere (starting in 2014) then reverses the situation, leaving the northern active latitudes more frequent connected to the solar wind and driving-up the mean rotation rate in the northern hemisphere. Balance is restored at the end of the declining phase of activity with a final pole-ward surge in the northern hemisphere (starting in 2016) returning the source-latitude distribution to a near-dipolar configuration. This sequence shows that changes in the distribution of northern and southern active regions can drive asymmetry in the resulting coronal rotation rate. 

Examining more closely the PFSS models shown in Figure \ref{figure_3Dplot}, the azimuthally averaged rotation rate of the open magnetic field is plotted in Figure \ref{figure_1_AZAV} along with 1D cuts of rotation at various radial distances (yellow being the source surface, and darker colours moving down towards the solar surface). Grey regions indicate completely closed latitudes, i.e when field lines are traced from all longitudes at this latitude, they are all closed therefore there is no value to return. The snapshots from activity minima show rotation profiles that are roughly north-south symmetric, with the exception of the location of the closed field in grey. These cases are dominated by the slowly rotating polar sources, with some faster equatorial connectivity associated with small active regions (visible in Figure \ref{figure_3Dplot}). 

The models of rising (2nd panel) and declining (4th panel) activity both have a clear north-south asymmetry in rotation, with the corona rotating systematically slower in the southern hemisphere. As discussed, this is due to the imbalance of source regions between the two hemispheres. This is clear from Figure \ref{figure_latitudes}, where the histogram shows a higher density of source regions in the low-latitude north than the south during these snapshots. This imbalance persists throughout most of the active phase of Cycle 24, except for during activity maximum in 2014 (in-between the pole-ward surges). Here the situation is reversed with more sources in the low-latitude south, leading to faster coronal rotation in the south. During this time, the dipole component of the Sun's magnetic field is weak or highly inclined, and so in Figure \ref{figure_1_AZAV} the closed regions (in grey) appear over the poles.

\subsection{Apparent rotation from coronal streamers}\label{lasco}

Given the wealth of coronal observations in scattered white light, recent works have begun to reconstruct the rotation of the corona based on the apparent motion of streamer structures. Most recently, \citet{edwards2022solar}, following the methodology of \citet{morgan2011rotation}, who measured the rotation rate of long-lived streamer structures in LASCO C2 white light images from 2008 - 2020. While there are many local deviations in the measured rotation rate from the surface rate, which may be of interest in the discussion of angular momentum transport in the low-corona. The mean coronal rotation rate from this method often deviates systematically from the surface rotation rate, with a flatter rotation profile that is closer to the Carrington rate (similar to that motivated in Appendix \ref{AP_CH}, see Figure \ref{figure_DRprofile}). However, these measurements come with challenges in interpretation, as streamers do not form at all latitudes during the solar cycle \citep[discussed in][]{morgan2011rotation}. At solar minimum streamers are confined to the equator, whereas during solar maximum streamers can be found at all latitudes (with challenges in accurately reconstructing their motion over the rotational poles). 

From the PFSS modelling performed in this study, the potential bias of using white light streamers to measure the overall rotation of the corona is assessed based on the available streamer latitudes during the cycle and the evolving rotation rate versus latitude of the corona. Figure \ref{figure_lasco} displays the azimuthally-averaged rotation rate at the source surface throughout the timeseries of magnetograms (panel a), along with the averaged scattered white light emission at three solar radii observed by LASCO C2 onboard the Solar and Heliospheric Observatory (panel b). The latitudinal variation of the mean coronal rotation and the streamer structures have an almost identical morphology. This is not unexpected as both quantities are driven by the underlying reconfiguration of the Sun's magnetic field. At solar minima, coronal rotation is dominated by the slowly rotating flows from the polar coronal holes and the streamers confined to the equator following the dipolar configuration of the large-scale magnetic field. With increasing activity the Sun's large-scale magnetic field becomes more complex, and so the sources of the solar wind move to the active latitudes and consequently faster rotating areas. The evolution of the Sun's large-scale magnetic field then allows for streamer structures that traverse a much broader range of latitudes.

The presence of white light streamers is essential for inferring coronal rotation at a given latitude and time in the cycle. From Figure \ref{figure_lasco} it is clear that streamers typically appear at latitudes with faster rotation rates. It might then be expected that the rotation profiles derived from white light streamers should systematically differ from the surface rotation profile. In our model, the rotation rate of coronal streamers is directly linked with the rotation rates at the source of the streamer structure. In which case, given that streamer structures are often anchored to the active latitudes, white light observations are more likely to produce a mean rotation rate that is flattened towards the Carrington rotation rate. This may explain some of the findings from these previous works \citep[i.e.][]{morgan2011rotation,edwards2022solar}.

\begin{figure*}
 \centering
  \includegraphics[trim=0cm 0cm 0cm 0cm, clip, width=\textwidth]{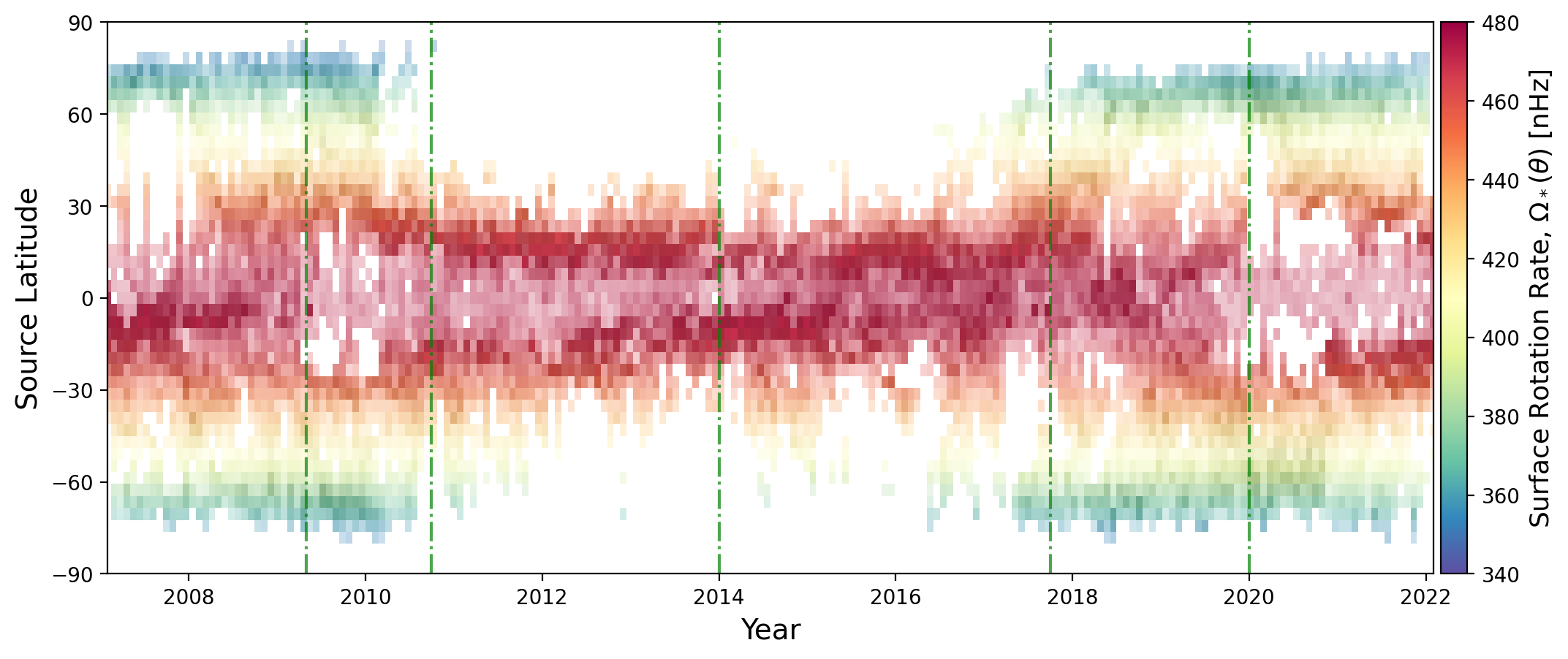}
   \caption{Same as the lower panel of Figure \ref{figure_latitudes}, but now field lines are traced down from the source surface only around the equator ($\pm15$ degrees). The histogram density is coloured by surface rotation rate.} 
   \label{figure_latitudesEquator}
\end{figure*}

\subsection{Equatorial connectivity and rotation rate}

With the exception of the \textit{Ulysses} spacecraft \citep[an overview of observations is presented in][]{mccomas2008weaker}, contemporary in-situ measurements of the solar wind are limited to the ecliptic plane of the solar system (this will change in future as ESA/NASA's Solar Orbiter mission begins higher inclination orbits in 2025). When trying to connect the theoretical results presented here to in-situ measurements, it is important to consider the sources of equatorial solar wind and their associated rotation rates. Figure \ref{figure_latitudesEquator} displays a similar analysis to that of Figure \ref{figure_latitudes}, but now with only field lines from $\pm 15$ degrees of the equator at the source surface being traced down to their sources. The histogram is coloured by the rotation rate at the surface, assuming the rotation profile from \citet{snodgrass1983magnetic}. 

Low-latitude sources of the solar wind are present throughout the entire solar cycle. In-situ measurements may then be expected to find slightly larger tangential speeds (no more than a few km/s). The tangential speeds measured in-situ depend strongly on the magnetic stresses that support the rotation of coronal plasma out to larger distances than the source surface (the tangential speed is equivalent to the rotation rate multiplied by the radial distance). In this regard, most of these low-latitude sources will have strong magnetic fields, and so these features could play a more significant role than indicated by the PFSS modelling in this study, enforcing their rotation higher up in the solar corona. In which case, the tangential speed of the solar wind may be more strongly influenced (rigid-rotation up to $1.5R_{\odot}$ results in a $\sim 7\%$ increase in the Sun's wind-braking torque). Measurements of the middle-corona are needed in order to evaluate impact of strong active regions on coronal rotation \citep[investigations into this area have become more frequent][]{west2022review, Chitta2022}.

Large tangential solar wind speeds are frequently measured by Parker Solar Probe in the near-Sun environment \citep[up to 50km/s][]{kasper2019alfvenic}, which indicates that angular momentum transport in the corona is more complicated than MHD models predicts \citep[the expectation is nearer to 5-10km/s;][]{reville2020role}. \citet{finley2020solar} found that these tangential flows could be made consistent with the expected angular momentum-loss rate of the Sun if averaged together with regions of slow and retrograde rotation, also detected in the equatorial solar wind by Parker Solar Probe. The model presented here does not allow for absolute retrograde rotation in the corona, given that the surface rotation profile is always prograde. Stream interactions between fast and slow solar wind sources are frequently employed to explain the existence of strong positive and negative tangential flow deflections in the solar wind at 1au \citep{yermolaev2018dynamics}. However, with Parker Solar Probe observing these flows so close to the Sun, it seems unlikely that wind-interactions will have had enough time to develop. To explain this, a strong contrast in the effective rotation rate between fast and slow solar wind streams (anchored at different latitudes) may be required to increase the frequency of collisions during the wind acceleration process.

\subsection{Potential impacts for other Sun-like stars}
In this study, the Sun's current differential rotation has been shown to have little influence on its wind-braking torque when averaged over a solar cycle. This results from the configuration of the Sun's magnetic field and its relatively weak differential rotation profile. During the lifetime of the Sun, however, differential rotation could have had a larger impact, depending on its initial rotation rate and subsequent degree of differential rotation thereafter. In both observations and numerical simulation, the amplitude of differential rotation $\Delta \Omega_*$ scales roughly with rotation rate $\Omega_*$ to a power $n$ \citep[e.g.][]{barnes2005dependence}. Taking $n=0.46$ \citep[from the global MHD simulations of][]{brun2022powering}, and assuming a self-similar solar differential rotation profile, for a young rapidly rotating Sun-like star with $\Omega_*/\Omega_{\odot}=5$ ($\Omega_*$ = 2278 nHz), the differential rotation contrast would be $5^{0.46} = 2.1$ times stronger than that of the current Sun ($\Delta\Omega_*\approx 265 nHz$). With a similar variation in wind source latitudes as the Sun, the value of $\langle \Omega_{wind}\rangle$ would vary from $\sim 1842$ nHz to $\sim 2326$ nHz, meaning the wind-braking torque could decrease from the solid body value by $20\%$ during a solar minimum-like field configuration or increase by $2\%$ during a solar maximum-like configuration. 

The scaling of stellar magnetic field strengths, and their large-scale magnetic field components continues to be constrained by Zeeman-Doppler imaging \citep{vidotto2014stellar, see2019non}. Given that very young Sun-like stars tend to possess stronger, and more dipolar field configurations, $\langle \Omega_{wind}\rangle$ may be systematically lower than $\Omega_*$ during the early main sequence. This can be further investigated by examining the differential rotation self-consistently produced in 3D MHD simulations of stellar interiors \citep[e.g.][]{brun2022powering}. More precision than this requires additional information on the heating and structuring of the stellar coronae, which is needed to determine the latitudinal distribution of the stellar wind sources, and thus the true impact of surface differential rotation. 

For other low-mass stars (0.2 to 1.3 solar masses), it is possible that the source latitudes of their stellar winds could produce effective rotation rates that differ significantly from the rotation period of the star recovered from their spot-modulated light curves. Asteroseismic inversions from \citet{benomar2018asteroseismic} recover latitudinal differential rotation profiles for Sun-like stars which are much stronger than the solar case. The mean amplitude of the differential rotation in their sample is six times stronger than that of the current Sun, which means that variation of the stellar wind source latitudes would have a much more pronounced effect on the wind-braking torque. Interestingly, the strength of differential rotation has been observed to vary with spectral type \citep{reiners2006rotation}, and in some cases shown to be significantly enhanced in F-type stars \citep{marsden2006surface}. This may lead to a systematic divergence between physics-based rotation-evolution models and the observed rotation period distributions. 

In old, slowly rotating main sequence stars, differential rotation may have an interesting effect during the (as of yet undetected) transition to anti-solar differential rotation \citep[the search for anti-solar rotators is detailed in][]{noraz2022hunting}. As the stellar magnetic field weakens and potentially loses its cyclic nature \citep{brun2022powering, noraz2022impact, kapyla2022transition}, the footpoints of the resulting axisymmetric dipole would be confined to the more rapidly rotating poles, resisting the decrease in wind-braking torque with age. This would challenge the current hypothesis of weakened magnetic braking \citep{van2016weakened}, with angular momentum being more efficiently lost due to the favourable configuration of the stellar magnetic field and differential rotation pattern. It seems more likely that something prevents Sun-like stars from entering this configuration, or that a decreasing mass-loss rate could counteract this effect, producing the expected weakening of the wind-braking torque \citep[discussed in][]{metcalfe2022origin}.

The recent study of \citet{tokuno2022transition} investigates the influence of differential rotation on the rotation-evolution of Sun-like stars. Their model allows for the amplitude of the differential rotation at the stellar surface to evolve in time, as a function of the Rossby number (rotation period normalised by the convective turnover timescale). The effective rotation rate used in the wind-braking torque of \citet{matt2015mass} is then taken from a low-latitude region, with the authors finding a weakening of the wind-braking torque at late-ages as this rotation rate is smaller than the underlying solid-body value. This study shows that the effective rotation rate used in the wind-braking torque depends on the dominant source latitudes of the stellar wind (during the wind-braking timescale), which is largely governed by the stellar magnetic field.

\section{Conclusion}
This study provides evidence to suggest that the current Sun's mean angular momentum-loss rate is not strongly influenced by the observed surface differential rotation (with respect to adopting a solid-body rotation rate). The equatorial solar wind, in which the majority of the solar wind angular momentum flux is transported, is supplied by a range of low-latitude sources, along with flows from the equator-ward edges of polar coronal holes. Thus the effective rotation rate of the solar wind remains close to the Carrington rotation rate during most of solar cycle, except during solar minima. At these times, when there are only a few small low-latitude source regions, the mean rotation rate of the corona decreases by 50-60 nHz with respect to the Carrington rate. This coincides with the cyclic minimum of the Sun's angular momentum-loss rate, and so the net impact, when averaged over the solar cycle, is strongly limited. 

Differential rotation could have a stronger influence on the wind-braking of other Sun-like stars with larger contrasts in rotation between their equator and poles (typically observed in the younger rapidly rotating stars). The degree to which this differential rotation will impact their wind-braking torque is dependent on the long-term latitudinal distribution of their stellar wind sources, which remains uncertain. This may change in future when better observational constraints on the latitudinal distribution of starspots throughout the main-sequence of Sun-like stars become available \citep[e.g.][]{berdyugina2005starspots, shapiro2014variability, morris2017starspots, icsik2018forward}. It is left for future work to ascertain the importance of differential rotation on the wind-braking torque on evolutionary timescales.

The PFSS modelling adopted in this work produces a static, force-free model of the corona. This is a rapid and computationally inexpensive method for assessing the likely variation of connectivity throughout the time-series of magnetograms used in this work \citep[see also the work of][]{badman2020magnetic, stansby2021active}. Our study assumes a direct relation between the observed surface rotation rate and that of the solar wind above, however many open questions still surround the rotation of the corona. It is likely that the act of differential rotation on the overlying coronal magnetic field will generate currents that modify the balance of forces in the corona. This effect can be found in the magnetofrictional models of \citet{yeates2008modelling}, who utilise a time-varying photospheric magnetic field boundary condition with a finite magnetic field relaxation timescale \citep[see also][]{van2000mean, hoeksema2020coronal}. This allows for complexity to develop in the corona, which is otherwise missing in force-free models. Observed coronal features are often better matched by this kind of modelling \citep{meyer2020investigation}. Magnetofrictional modelling has also been applied to other Sun-like stars \citep[see][]{gibb2016stellar}. Any potential hysteresis of the coronal magnetic field will likely change the source latitudes of the solar wind, and the degree of which the magnetic field enforces the surface rotation rate. 

Coronal rotation has impacts in many areas of active research, such as the accuracy of ballistic back-mapping of the solar wind when identifying photospheric sources \citep[e.g.][]{macneil2022statistical}, the production of accurate models of the inner heliosphere, and the overall forecasting of space weather. Thus, in the coming decade, studies of coronal rotation ranging from the distortion of coronal hole boundaries, up to the variation in white light streamers, and in-situ measurements of solar wind deflections, will be required to understand the evolution of angular momentum from the solar surface out into the solar wind.

\begin{acknowledgements}
This research has received funding from the European Research Council (ERC) under the European Union’s Horizon 2020 research and innovation programme (grant agreement No 810218 WHOLESUN), in addition to funding by the Centre National d'Etudes Spatiales (CNES) Solar Orbiter, and the Institut National des Sciences de l'Univers (INSU) via the Programme National Soleil-Terre (PNST).
This work utilizes data produced collaboratively between Air Force Research Laboratory (AFRL) and the National Solar Observatory (NSO). The ADAPT model development is supported by AFRL. The input data utilized by ADAPT is obtained by NSO/NISP (NSO Integrated Synoptic Program). NSO is operated by the Association of Universities for Research in Astronomy (AURA), Inc., under a cooperative agreement with the National Science Foundation (NSF).
The sunspot number used in this work are from WDC-SILSO, Royal Observatory of Belgium, Brussels. 
Data supplied courtesy of the SDO/HMI and SDO/AIA consortia. SDO is the first mission to be launched for NASA's Living With a Star (LWS) Program.
The SOHO/LASCO data used here are produced by a consortium of the Naval Research Laboratory (USA), Max-Planck-Institut fuer Aeronomie (Germany), Laboratoire d'Astronomie (France), and the University of Birmingham (UK). SOHO is a project of international cooperation between ESA and NASA.
Data manipulation was performed using the numpy \citep{2020NumPy-Array}, scipy \citep{2020SciPy-NMeth}, and pySHTOOLS \citep{wieczorek2018shtools} python packages.
Figures in this work are produced using the python packages matplotlib \citep{hunter2007matplotlib}, and Mayavi \citep{ramachandran2011mayavi}.
\end{acknowledgements}

%
%

\bibliographystyle{yahapj}
\bibliography{adam}

\begin{appendix}
\section{Fitting the differential rotation of coronal holes}\label{AP_CH}

\begin{figure*}
 \centering
  \includegraphics[trim=0cm 0cm 0cm 0cm, clip, width=\textwidth]{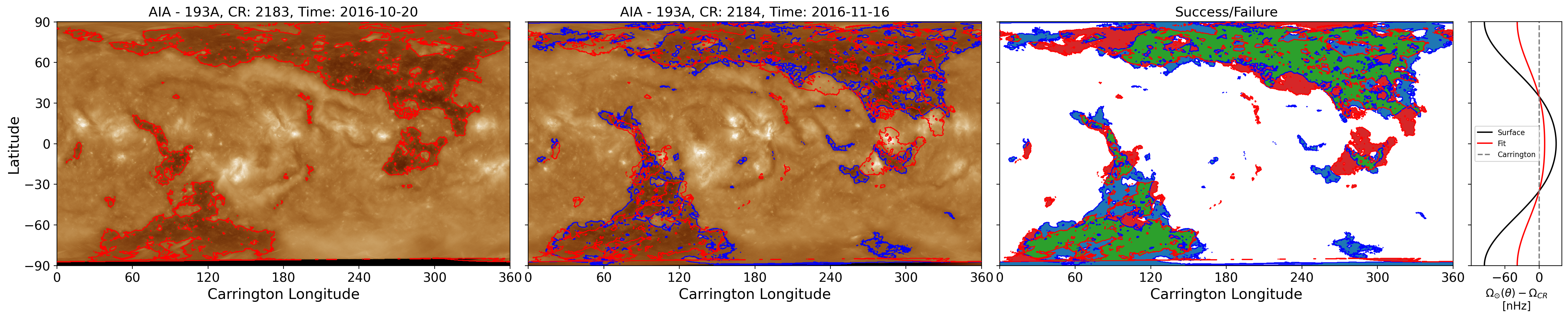}
  \includegraphics[trim=0cm 0cm 0cm 0cm, clip, width=\textwidth]{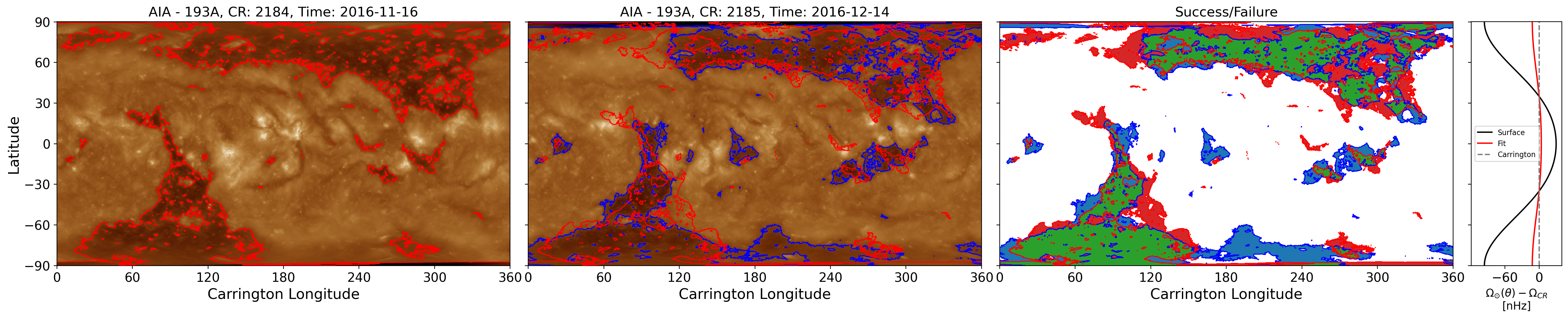}
  \includegraphics[trim=0cm 0cm 0cm 0cm, clip, width=\textwidth]{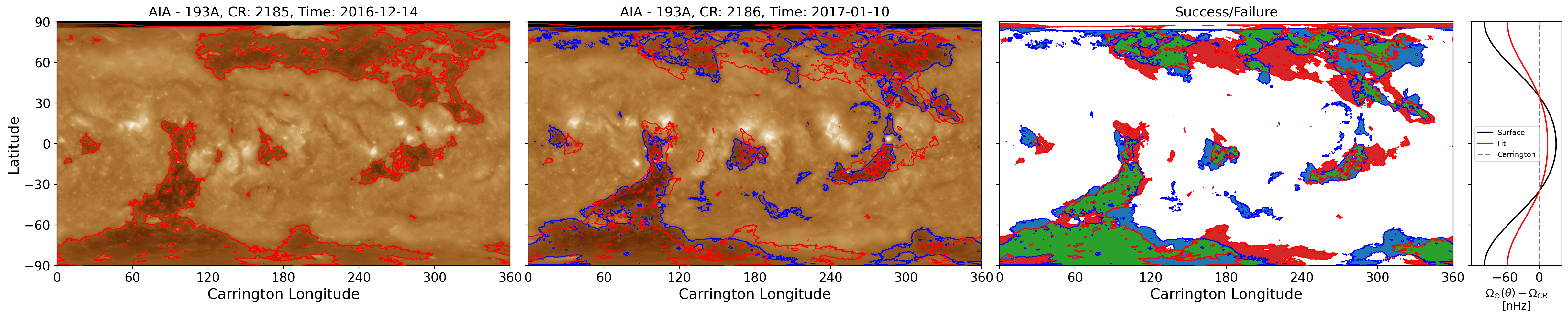}
   \caption{Fitting the differential rotation profile of the coronal holes in SDO/AIA-193A images during CR2183-CR2186. Each row repeats the following sequence. The left column depicts a given synoptic EUV chart in the sequence CR2183 - CR2185 with a red contour identifying the coronal hole area using a brightness threshold. The second column shows the next synoptic chart (CR2184 - CR2186) with a blue contour identifying the coronal hole area in the same manner as the leftmost column. The area inside these contours is assumed to be representative of the coronal hole area for each synoptic chart. The red contour from the first column is then distorted using equation (\ref{long_shift}) by varying the $\Omega_{eq}$, $\alpha_2$, and $\alpha_4$ parameters, in order to find a maximum overlap of the area between the two contours. The `best fit' distorted contour is over-plotted in the second column in red. The third column depicts the quality of the overlap between the red distorted coronal hole area and the blue coronal hole area, with green representing a match and red/blue indicating that there is area in the red/blue contour that does not match. The final column shows the surface rotation rate versus latitude using in this study in black, with the best fit rotation profile in red. The average fit parameters from all three retrievals are $\Omega_{eq}$= 463.0 nHz, $\alpha_2$= -26.7 nHz, and $\alpha_4$= -33.1 nHz.}
   \label{figure_A1}
\end{figure*}

In order to provide an illustrative rotation profile to contrast with the well-known surface differential rotation pattern, the evolution of two trans-equatorial coronal holes in 2016 - 2017 are examined. Synoptic AIA 193$\AA$ charts from Carrington Rotations (CRs) 2183, 2184, and 2185 are shown in the leftmost column of Figure \ref{figure_A1}. The coronal hole boundaries are highlighted in red by selecting a threshold on the brightness of the EUV emission (around the 30th percentile). This process is repeated in the center column for CR 2184, 2185, and 2186, with the same threshold brightness shown in blue. An algorithm is employed to distort the red contours from the leftmost column following the form of solar-like differential rotation described by equation (\ref{omega_*}). The longitudinal shift applied to the contour follows the form,
\begin{eqnarray}
    \delta \phi_{CR}(\theta) &=& t_{CR} [\Omega_*(\theta)-\Omega_{CR}],\\
    &=& t_{CR} \bigg[ \Omega_{eq} + \alpha_2\cos^2\theta+\alpha_4\cos^4\theta -\frac{360^{\circ}}{t_{CR}} \bigg].
    \label{long_shift}
\end{eqnarray}
where $t_{CR}=25.4$ days is the Carrington rotation period, with the values of $\Omega_{eq}$, $\alpha_2$, and $\alpha_4$ are each allowed to vary in an attempt to match the coronal hole boundaries from the following CR. The algorithm attempts to maximise the area of agreement between the two contours. The best fit distorted contour in each case is over-plotted in the centre column in red. The two right panels show the agreement or disagreement between the area inside the contours, along with the fit rotation profile in comparison to the typical surface differential rotation rate. Agreement between contours is filled in green, area where the coronal hole exists in the distorted coronal hole but not in the chart attempting to be fit is filled in red, and the contrary in blue.

The observed trans-equatorial coronal holes in EUV are clearly distorted during the course of four solar rotations, however this appears to be less than would be predicted by the typical surface differential rotation profile. In particular, there are a few equatorial coronal holes in this timeseries that remain at an almost fixed longitude, i.e. rotating at the Carrington rotation rate. Thus the rotation rate at base of the solar wind may not correspond exactly to that of the photosphere. The mean fit parameters from Figure \ref{figure_A1} are used as an illustrative example for comparison with the typical surface differential rotation rate throughout this work.

\section{Influence of the Source Surface Radius}\label{AP_RSS}

\begin{figure*}
 \centering
  \includegraphics[trim=0cm 0cm 0cm 0cm, clip, width=\textwidth]{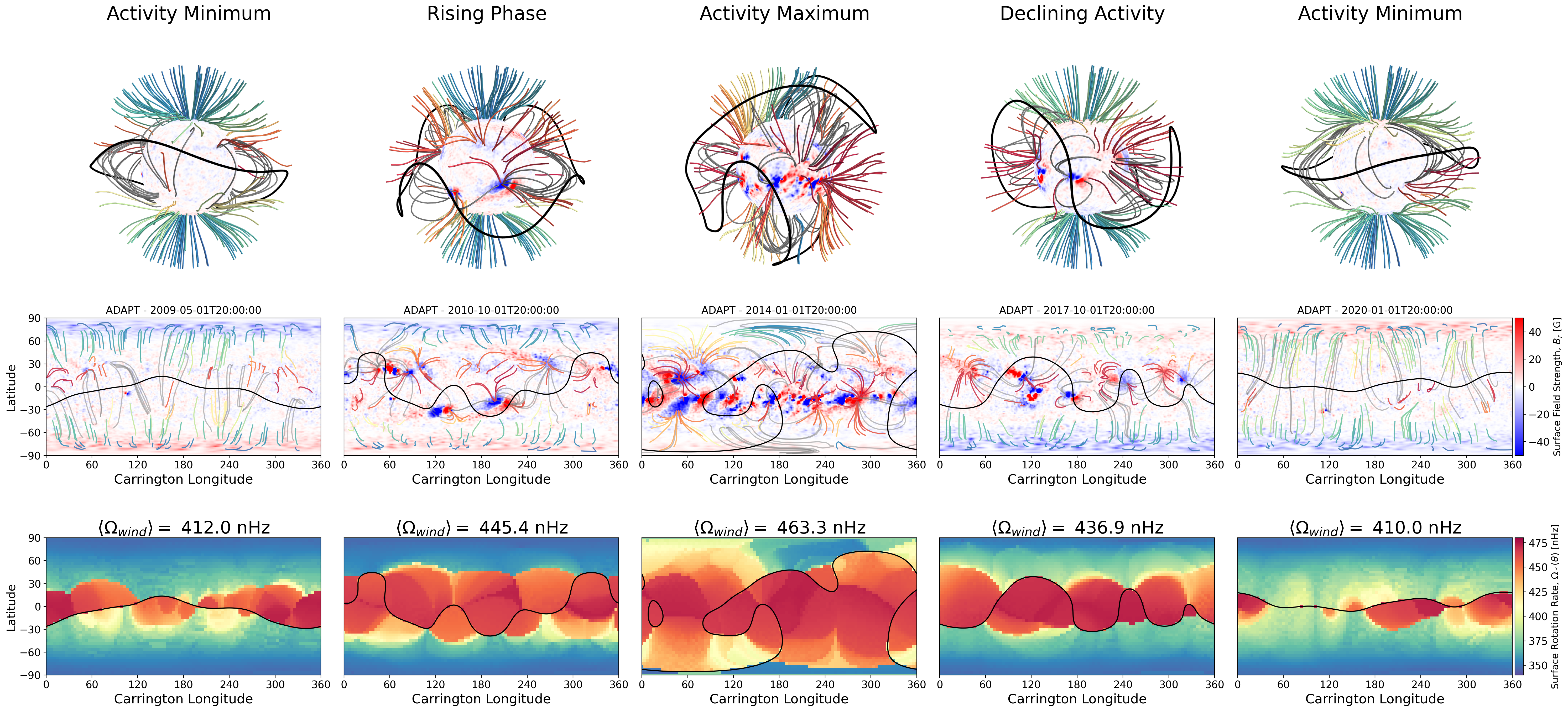}
   \caption{Same as Figure \ref{figure_3Dplot}, now with a source surface radius of 2.0 $R_{\odot}$.}
   \label{figure_3Dplot_rss2}
\end{figure*}

\begin{figure*}
 \centering
  \includegraphics[trim=0cm 0cm 0cm 0cm, clip, width=\textwidth]{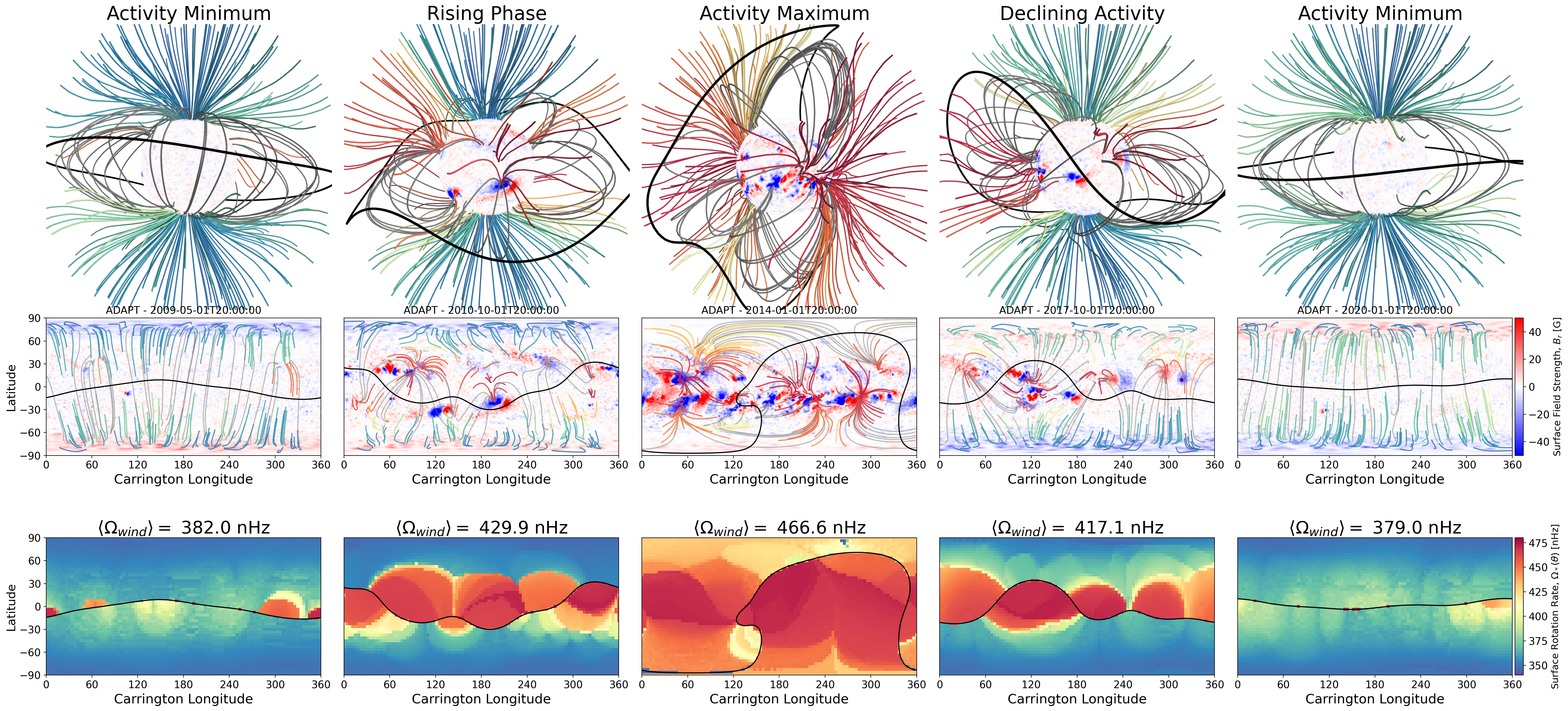}
   \caption{Same as Figure \ref{figure_3Dplot}, now with a source surface radius of 3.0 $R_{\odot}$.}
   \label{figure_3Dplot_rss3}
\end{figure*}

\begin{table*}
\caption{Effective Rotation Rates with Varying Source Surface Radius (the Carrington rate is $\sim 456$ nHz).}
\label{table_rotation2}
\centering
\begin{tabular}{c c| c c c } 
\hline\hline
Phase & Date  & $\langle \Omega_{wind}\rangle$ [nHz]& & \\
 & & $R_{ss}=2.0R_{\odot}$ & $R_{ss}=2.5R_{\odot}$ & $R_{ss}=3.0R_{\odot}$ \\
\hline
   Activity Minimum &2009-05-01 & 412.0 & 392.5 & 382.0 \\
   Rising Phase &2010-10-01 & 445.4 & 437.8 & 429.9 \\
   Activity Maximum &2014-01-01 & 463.3 & 464.7 & 466.6 \\
   Declining Activity &2017-10-01 & 436.0 & 426.0 & 417.1 \\
   Activity Minimum &2020-01-01 & 410.0 & 387.9 & 379.0 \\
\hline
\end{tabular}
\end{table*}

The source surface radius $R_{ss}$ sets the distance at which the coronal magnetic field becomes purely radial in the PFSS model. In general, reducing the size of the source surface allows more magnetic field to open into the solar wind which comes from smaller scale magnetic features at the photosphere. Increasing the source surface radius allows for larger closed magnetic field loops and so the open magnetic field at the source surface is reduced. In this study, for the PFSS modelling the source surface radius is fixed to $2.5 R_{\odot}$, which is a typical value from the literature. However, it is known that the size of the source surface radius often needs to be changed in order to match in-situ observations \citep{panasenco2020exploring,badman2020magnetic}, or observations of the coronal hole areas \citep{linker2017open}, and even when making direct comparisons between PFSS and MHD models \citep{reville2015solar}. To test the dependence of our result on our chosen source surface radius, the calculations from Section \ref{results} are performed again for the source surface radii of 2 and 3 $R_{\odot}$. Figures \ref{figure_3Dplot_rss2} and \ref{figure_3Dplot_rss3} contrast the difference in connectivity of the magnetic field models. The resulting values of $\langle \Omega_{wind}\rangle$ for each of these epochs with varying source surface radii are compiled in Table \ref{table_rotation2}. Notably, during solar minimum increasing the source surface radius closes off some of the low-latitude source of the solar wind found in the model with $R_{ss}=2.5R_{\odot}$ in Figure \ref{figure_3Dplot}, whereas decreasing the source surface radius makes these regions more dominant in the calculation of the effective rotation rate. 
\begin{figure}
 \centering
  \includegraphics[trim=0cm 0cm 0cm 0cm, clip, width=0.49\textwidth]{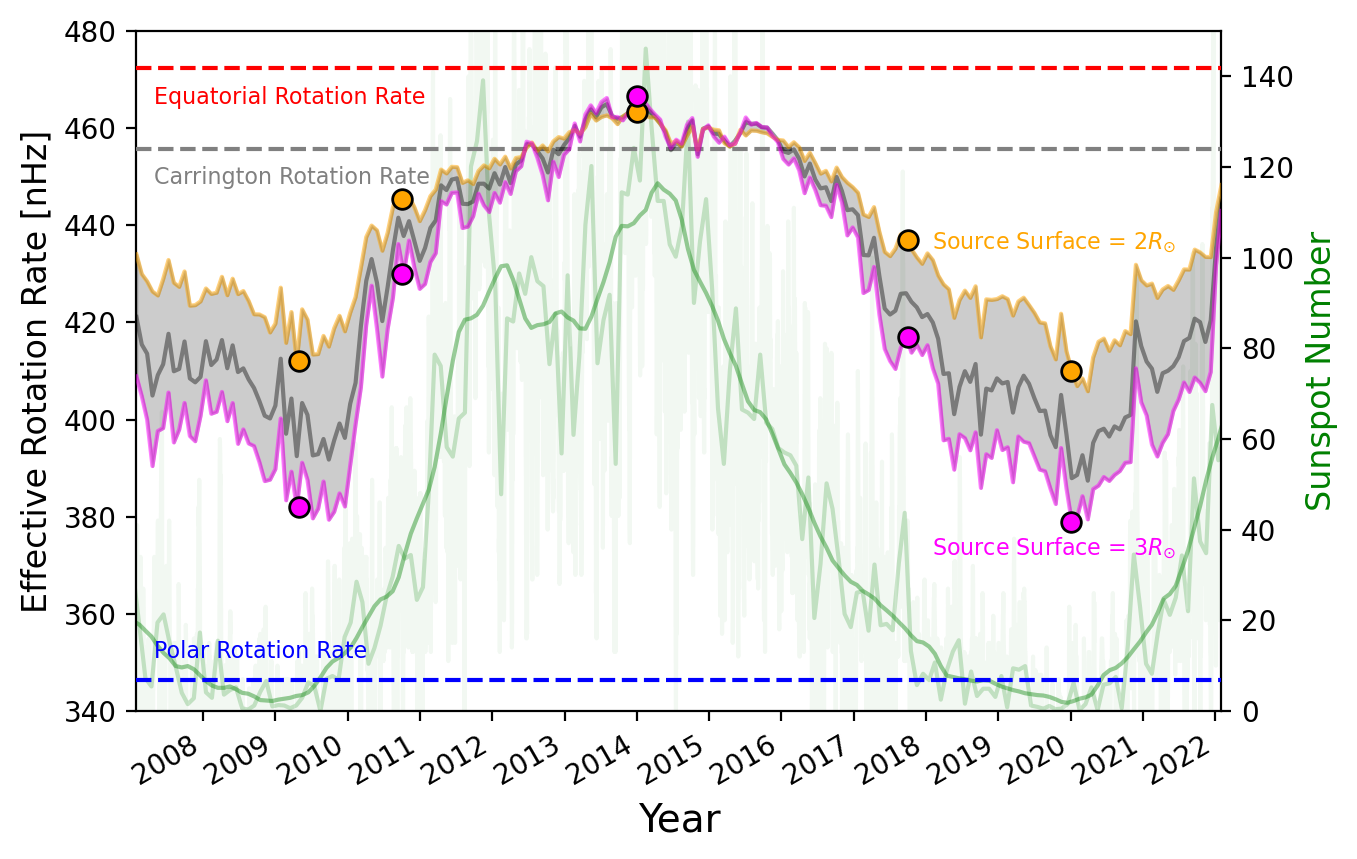}
   \caption{Same as Figure \ref{figure_effRotation}, now with showing the influence of varying the PFSS source surface radius from 2 - 3 $R_{\odot}$. The PFSS models shown in Figures \ref{figure_3Dplot_rss2}, and \ref{figure_3Dplot_rss3} are highlighted.}
   \label{figure_effRotation_rss}
\end{figure}

The variation in the effective rotation rate of the solar wind between the two extreme source surface radii is explored in Figure \ref{figure_effRotation_rss}. There is very little difference between the result from the two models during solar maximum, due to the more effectively-homogeneous distribution of photospheric sources in latitude. Here the larger source surface model can produce a slightly larger effective rotation rate, which is a result of closed field over the rotation poles being opened which has a slower rotation rate. The largest differences come at solar minima where the source latitudes of the solar wind is highly sensitive to the size of the source surface. The smaller source surface connecting to more low-latitude sources of the solar wind and so have a larger effective rotation rate, and vice versa.

\end{appendix}

\end{document}